\documentclass[prb,aps,twocolumn,superscriptaddress,showpacs]{revtex4}
\usepackage{graphicx}
\usepackage{psfrag}
\usepackage{color}
\usepackage{subfigure}
\usepackage{amsmath}
\definecolor{dred}{rgb}{0.7,0.0,0.0}

\begin{document}

\title{Properties of a Two Orbital 
Model for Oxypnictide Superconductors: \\
 Magnetic Order, $B_{\rm 2g}$ Spin-Singlet Pairing Channel, and Its Nodal Structure. }

\author{A. Moreo}
\affiliation{Department of Physics and Astronomy, The University of Tennessee, Knoxville, TN
 37996}
\affiliation{Materials Science and Technology Division, Oak Ridge National Laboratory, Oak Ridge, TN 32831}

\author{M. Daghofer}
\affiliation{Department of Physics and Astronomy, The University of Tennessee, Knoxville, TN
 37996}
\affiliation{Materials Science and Technology Division, Oak Ridge National Laboratory, Oak Ridge, TN 32831}

\author{J. A. Riera}
\affiliation{Instituto de F\'isica Rosario, Consejo Nacional de Investigaciones Cient\'ificas y T\'ecnicas, Universidad Nacional de Rosario, 2000-Rosario, Argentina}

\author{E. Dagotto}
\affiliation{Department of Physics and Astronomy, The University of Tennessee, Knoxville, TN
 37996}
\affiliation{Materials Science and Technology Division, Oak Ridge National Laboratory, Oak Ridge, TN 32831}

\date{\today}

\begin{abstract}
A recently proposed two orbital model for the new Fe-based superconductors is studied
using the Lanczos method on small clusters as well as pairing
mean-field approximations. Our main goals are {\it (i)} to provide  
a comprehensive analysis of this model using numerical techniques 
with focus on the magnetic state at half-filling and
the quantum numbers of the state with two more electrons than
half-filling
and {\it (ii)} to investigate the nodal structure of the mean-field
superconducting state and compare the results with angle-resolved
photoemission data. In particular, we provide evidence that the dominant 
magnetic state at half-filling contains spin ``stripes'', as observed 
experimentally using neutron scattering techniques. Competing spin states are 
also investigated. 
The symmetry properties of the state with two more electrons added to half 
filling are also studied: depending on parameters, either a spin singlet or
spin triplet state is obtained. Since experiments suggest spin singlet pairs, 
our focus is on this state. Under rotations, the spin-singlet state transforms 
as the $B_{\rm 2g}$ representation of the $D_{\rm 4h}$ group.
We also show that the $s\pm$ pairing operator transforms according to
the $A_{\rm 1g}$ representation of $D_{\rm 4h}$ and becomes dominant only in
an unphysical regime of the model where the undoped state is an
insulator. We obtain qualitatively very similar results both with hopping
amplitudes derived from a Slater-Koster approximation and with
hoppings selected to fit band-structure calculations, the main 
difference between the two being the size of the Fermi surface pockets. 
For robust values of the effective electronic attraction
producing the Cooper pairs, assumption compatible with recent
angle-resolved photoemission (ARPES)  results that suggest a small
Cooper-pair size, the nodes of the two-orbital model  are found to be
located only at the electron pockets.  
Note that recent ARPES efforts have searched for nodes at the hole
pockets or only in a few directions at the electron pockets.
Thus, our results for the nodal distribution will help to guide future ARPES
experiments in their search for the existence of nodes in the new Fe-based superconductors.   
More in general, the investigations reported here aim to  establish several 
of the properties of the two orbital model. Only a detailed comparison with
experiments will clarify whether this simple model is or not a good approximation to describe the
Fe pnictides.
\end{abstract}


\maketitle

\section{Introduction}

\subsubsection{Current status of experimental and theoretical investigations}

The discovery of a new family of superconducting materials with Fe-As
layers in their
structure\cite{Fe-SC,chen1,chen2,wen,chen3,ren1,55,ren2} has
triggered a large effort in the condensed matter community.  
$\rm LaO_{\it 1-x} F_{\it x} Fe As$ is a much studied example of this family
of compounds. The $\sim 55$~K record critical temperature\cite{55} in
$\rm SmO_{\it 1-x} F_{\it x} Fe As$
is second only to those observed in the Cu-oxide family of high temperature
superconductors. In addition, there are several aspects of the physics of the new
Fe-based superconductors that suggest the possibility of an exotic pairing
mechanism at work:

{ (1)} Evidence is accumulating that phonons may not be sufficient
to understand the superconductivity of these 
compounds.\cite{phonon0,phonon1,phonon2} Moreover, the importance of correlations between the
electrons has been remarked in several
investigations.\cite{sefat,liu,haule2,haule,dubroka,boris,C.Liu,J.Zhao,kohama,imada,H.Liu}
In fact, it has been claimed that 
these oxypnictide superconductors may bridge the gap between MgB$_2$
and the  Cu-oxide superconductors.\cite{Jaro,basov} 
In addition, a pseudogap was detected, similarly as in the
cuprates.\cite{Y.Ishida,T.Sato,liu2,L.Zhao} Coexistence or proximity
of magnetism and superconductivity has also been reported.\cite{A.Drew,I.Felner,takeshita,H.Chen}
Although the parent undoped compound
is not a Mott insulator, these results suggest that the influence of electron-electron repulsions cannot be neglected.
Perhaps the {\it intermediate} range of ``$U/t$'', where $U$ is the 
typical Hubbard repulsion scale and $t$ the
typical hopping amplitude in a tight-binding description, is the most representative of the new superconductors.
$U$ cannot be too large, otherwise the system would develop a gap and the undoped compound would be insulating, contrary to the experimentally observed properties of the undoped limit
that suggest bad metallic behavior. 
But poor-metal characteristics imply that
$U$ cannot be too small either, otherwise the undoped system 
would be a good metal. In addition, the mere presence of a spin-density-wave
magnetic state shows that correlations must be important.

{ (2)} Several experimental investigations suggest the presence of 
nodes in the superconducting
gap.\cite{nodal1,nodal2,nodal3,Ahilan,nakai,Grafe,Y.Wang,matano,mukuda,millo,wang-nodes}
This is reminiscent of the nodes that appear in the $d$-wave superconducting state of the high-Tc
cuprates.  However, other investigations indicate nodeless 
superconductivity.\cite{hashimoto,arpes,arpes2,C.Martin,T.Chen,parker,mu} As a consequence, this issue is still controversial.

{ (3)} 
The undoped parent compound has long-range spin order
in the ground state.\cite{sdw} 
This magnetic state corresponds to spin ``stripes'' having the Fe spins along one of the Fe-Fe crystal
axes pointing all in the same direction, and being antiferromagnetically coupled in the perpendicular direction.  
According to neutron scattering experiments, 
in LaOFeAs the transition to this magnetic state occurs at 134~K, and the magnetic
moment is 0.36~$\mu_{\rm B}$, which is smaller than anticipated.\cite{neutrons1} 
For NdOFeAs,\cite{neutrons2} the critical temperature is 141~K
and the magnetic moment is even smaller 0.25~$\mu_{\rm B}$. However, 
recently by means of resistivity, specific heat, and magnetic susceptibility measurements,
the antiferromagnetic critical temperature of SrFe$_2$As$_2$ was reported to be as high as 205~K, with a more robust Fe magnetic
moment of value 1.7~$\mu_{\rm B}$.\cite{neutrons3} Also, CaFe$_2$As$_2$ was
investigated using neutron diffraction, and 
a critical temperature 173~K with a moment 0.8~$\mu_{\rm B}$ was reported.\cite{neutrons4}
Thus, although originally it was believed that the undoped material had a very weak magnetic state,
the most recent results suggest that the spin striped order may be more robust.

On the theory front, several band-structure calculations have shown that the
Fermi surface of these and related compounds is made out of
two small hole pockets centered at the $\Gamma$ point, 
and small electron pockets at the $X$ and $Y$ points, in the
notation corresponding to a square lattice of Fe atoms.\cite{first,singh,xu,cao,fang2}
These calculations have also shown that the 3$d$ levels of Fe play the dominant
role in establishing the properties of these materials near the Fermi level.
To address theoretically
the physics of these compounds, particularly the superconducting state, 
model Hamiltonians are needed and
several proposals for the dominant 
pairing tendencies have been made.\cite{kuroki,mazin,FCZhang,han,korshunov,baskaran,plee,yildirim,extra,yao,xu2,stratos}
In particular, a two orbital model based on the
$d_{\rm xz}$ and $d_{\rm yz}$ orbitals was recently presented.\cite{scalapino}
Several other investigations have addressed this model for the 
new superconductors, using 
a variety of 
approximations.\cite{li,qi,ours,hu,dhlee,zhou,lorenzana,sk,parish,choi,yang,calderon} 
Classifications of the possible superconducting order parameters for the two-orbital model 
have been made.\cite{wang,shi,2orbitals,wan} 
 
As already mentioned, a variety of experimental results 
suggest that the Cooper pairs are spin singlets.\cite{Grafe,matano,kawabata} 
Thus, it is important to find the
range of parameters leading to spin singlets 
in model Hamiltonians, 
since several 
calculations produce either 
singlet or triplet superconductivity depending on the couplings and bandwidths used. 
For this experimentally-based reason, our focus here will be mainly on singlet
superconductivity.

\subsubsection{Why the two orbital model?}

In this manuscript, a detailed study of the two orbital model for the oxypnictide superconductors
is carried out using Lanczos
and pairing mean-field techniques. This effort provides a comprehensive view of the model,
considerably expanding our recent research on the subject\cite{ours}
by varying the several couplings of the model and studying the main tendencies. When two electrons
are added to the half-filled ground state, a spin-singlet
state that  transforms in a non-trivial manner under rotations is shown to dominate in the regime of
couplings that is argued to be the most relevant 
to describe the new superconductors. In addition, the nodal structure
of the superconducting state obtained using these spin-singlet pairs is here studied 
for this model using the pairing
mean-field approximation. 
Our results are compared with recent ARPES experiments,
and suggestions to further refine the search for nodes in those experiments are discussed.

Currently there is no consensus on what is the minimal model capable of 
capturing the essential physics of the oxypnictides. Band structure calculations 
in the local-density approximation (LDA) indicate
that the bands that form the observed electron and hole pockets are strongly 
hybridized but they have mostly Fe-3$d$ character.\cite{vero,mazin} Several 
authors argue that the hybridization of the Fe-3$d$ is so strong that all 5 $d$ 
orbitals have to be considered to construct a minimal model. For instance, a five-orbital model has 
been proposed.\cite{kuroki} The tight-binding term respects the
FeAs lattice symmetries and the hopping parameters have been obtained from 
fittings against the LDA calculations. The parameters used reproduce the 
Fermi surface (FS) for the electron doped system (i.e. electronic density $n=6.1$) but an extra
hole pocket around M (in the notation of the extended 
Brillouin zone) appears
for the undoped case and upon hole doping. For this reason, the model may not 
be suitable to study the magnetic properties of the undoped system. 
In addition, the number of degrees of freedom in five-orbital models makes its study 
very difficult
using numerical techniques. However, LDA calculations have
shown that,
although heavily hybridized, the main character of the bands that determine 
the FS is $d_{\rm xz}$  and $d_{\rm yz}$, with a small contribution of $d_{\rm xy}$ at the 
most elongated portions of the electron pockets.\cite{fang2,vero} 
This fact has 
been the main justification for the proposal of two \cite{scalapino,ours} and three
\cite{plee} orbital models. The two-orbital model can have its hopping 
parameters fitted such that the shape of the FS, both in the undoped and 
electron and hole doped cases, are well reproduced in the reduced or folded BZ.
However, it has been argued by some authors \cite{plee} that the 
two hole-pockets around $\Gamma$ have to arise from the $d_{\rm xz}$ and $d_{\rm yz}$ 
orbitals that are degenerate with each other at $\Gamma$, as obtained in LDA. 
In the two-orbital model, one of the hole pockets forms around $M$ in the extended
BZ which gets mapped onto $\Gamma$ upon folding. The $d_{\rm xz}$ and $d_{\rm yz}$ 
orbitals that form the $M$-point pocket are degenerate at $M$ and, upon the folding, give rise to
higher energy bands at the $\Gamma$ point. For this reason 
one of the hole pockets in the two orbital model may not have 
the correct linear combination of orbitals, potentially leading to incorrect conclusions. In addition, 
it is  also argued that the contribution of the $d_{\rm xy}$ orbital to  
the electron pockets may play an important role that should not be ignored which motivated the proposal of the three-orbital model.\cite{plee} However, the three
orbital model cannot eliminate a spurious hole pocket around $M$. Thus, 
a fourth 
orbital needs to be added to accomplish this task and, again, the number of degrees 
of freedom makes this model too complex to be studied numerically. 

Then, the justification
for continuing studying a minimal model with just {\it two} orbitals,
 as carried out in the present 
manuscript, is the following: {\it (i)} The correct shape of the FS is reproduced in the 
{\it reduced} Brillouin zone, both in the doped and undoped cases.
{\it (ii)} The 
main character of all the bands that determine the FS is $d_{\rm xz}$ and $d_{\rm yz}$,
except for a small portion of the electron pockets that has $d_{\rm xy}$ 
character. Then, it it worthwhile to understand the role, if any, that this orbital 
plays in the magnetic and superconducting states. {\it (iii)} The two-orbital model is
the only one that can be studied exactly with numerical techniques using the minimal 
size cluster needed for a spin striped state.\cite{ours} Thus, we believe that it is very important to establish
which properties of the oxypnictides are properly captured by this model, and which ones not.
The role that the correct shape of the FS plays can be investigated as well, and also 
the pairing symmetry and nodal structure involving only the $d_{\rm xz}$ and $d_{\rm yz}$ orbitals. 
It is interesting to notice that although  two 
superconducting gaps may appear in a two orbital model,\cite{Suhl} symmetry 
forces the magnitude of the gaps to be the same in this case.\cite{wan}

\subsubsection{Organization}

The organization of the paper is as follows. In Sections II and III, the two-orbital model is derived. The emphasis
is on the Slater-Koster (SK) procedure to evaluate the hopping amplitudes, 
but the model derived by this method is more
general: it coincides with the two-orbital Hamiltonian proposed earlier,\cite{scalapino}
and the values of the hoppings can be
obtained also by fitting band-structure calculations.\cite{scalapino} 
Both sets of hopping  parameters will be
used in the following sections. 
The qualitative aspects of the magnetic and pairing states  are shown to be
the same for both sets of hopping amplitudes. In Section IV, results for the ground states of the undoped model
(half-filled) and the case of two more electrons than half-filling will be discussed using the Lanczos technique. 
The emphasis is on the dominant magnetic states and on 
the pairing tendencies, which are either in the spin
singlet or triplet channels depending on couplings. Moreover, the spin singlet case is shown to correspond to
the $B_{\rm 2g}$ representation of the $D_{\rm 4h}$ lattice symmetry group of the model. Section V contains a pairing
mean-field analysis of the nodal structure of the model. 
The two orbitals nature of the
problem causes the number and location of the nodes to be a more complex topic than
for just one orbital. A qualitative comparison with experiments is included here.
Section VI contains our main conclusions. The possible source of the
$B_{\rm 2g}$ pairing and the $s\pm$ pairing operator are discussed in the
appendices. 

\section{Model discussion and derivation of hopping amplitudes}

To study numerically the properties of $\rm LaO_{\it 1-x}F_{\it x}FeAs$ and related
compounds, it is necessary to construct a simple model, one that contains a
minimum amount of degrees of freedom but still preserves the main physics of the problem. 
Since all the materials
in the family have in common the  $\rm Fe-X$ planes (X=As, P, ...), as a first approximation
we will just focus on those planes, similarly as it occurs in theoretical studies of the 
Cu-O planes in the cuprate superconductors. In
addition, band structure calculations\cite{first,singh,xu,cao,fang2} have shown the relevance of
the Fe $3d$ levels, and that mainly two bands determine the Fermi surface (see Introduction). 
Based on these
considerations, here we will include only the $d_{\rm xz}$ and $d_{\rm yz}$ Fe orbitals in our discussion. To estimate
the hopping amplitudes for the tunneling from one Fe to another and, thus, define a tight-binding model,
we will calculate their  hybridization with the three $p$ orbitals of As
following the Slater-Koster formalism.\cite{slater} From the Fe-As hopping integrals,
we will calculate the effective Fe-Fe tight-binding hopping
parameters following a standard perturbative approach. 
Thus, the hopping parameters in this model will be
functions of the overlap integrals between the orbitals and the distance between the atoms. While this procedure is not
as accurate as band-structure calculations, it provides a simple to understand
approach that has ``ab-initio'' characteristics, can be easily reproduced 
since the calculations are analytical, and they also illustrate how the geometry of the
problem affects the hoppings. 

\begin{figure}[thbp]
\begin{center}
\includegraphics[width=8cm,clip,angle=0]{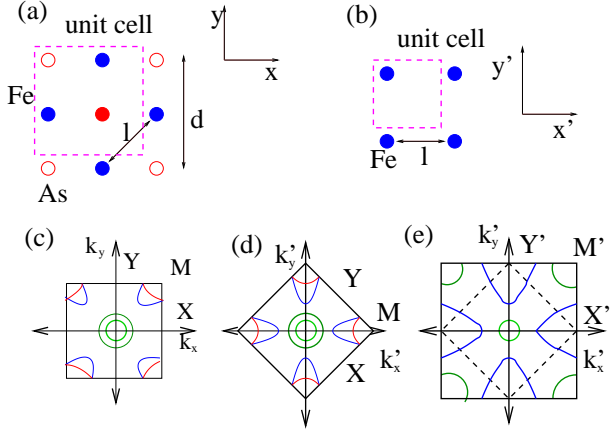}
\vskip 0.3cm
\caption{(Color online)  (a) Schematic representation of the Fe-As plane. 
Blue circles are the Fe
atoms. The red filled circle is an As atom at a distance $c$ below the plane, while the
red open circles  are As atoms at a distance $c$ above the plane. (b) Unit cell
for the effective Fe-only square lattice. The Fe-Fe lattice has been rotated by $45^o$.
(c) Schematic first Brillouin Zone (FBZ) for the Fe-As plane. The point $X$ is at $(2\pi/d,0)$, with
$d=\sqrt{2}l$. (d) FBZ for the Fe-As lattice after a $45^o$ rotation. (e) FBZ
for the rotated Fe-Fe shown in (b). $X'=(2\pi/l,0)$ and it is equivalent to the
$M$ point for the Fe-As plane in (c). The electron and
hole Fermi surfaces obtained by band-structure calculations are schematically indicated.
Panels c-e will be useful for the discussion related to the nodal structure 
of the superconducting state in Section V.
}
\label{Fig1}
\end{center}
\end{figure}

However, before proceeding, we remark that another avenue to obtain the hopping
amplitudes is via fittings of the band-structure calculations.\cite{scalapino} 
In our description of results below, data for both the SK hoppings 
and those that fit band structures will be presented. An important
result is that both sets of hoppings lead to similar
qualitative results, both in the undoped case, regarding the magnetic state, 
as for two electrons added,
regarding the pairing tendencies. 

The unit cell in the FeAs planes contains two Fe atoms, since the As atoms are above
and below the plane defined by the Fe atoms 
in alternating plaquettes (Fig.~\ref{Fig1}(a)). However,
after the calculation previously described  only the Fe atoms will be 
considered in a simple two-orbital Hamiltonian.
Since these Fe atoms form a planar square lattice, it is natural to orient
the lattice as in Fig.~\ref{Fig1}(b).


\begin{figure}[thbp]
\begin{center}
\includegraphics[width=8cm,clip,angle=0]{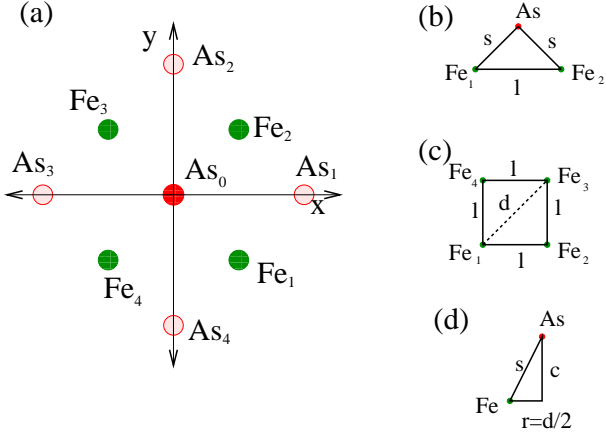}
\vskip 0.3cm
\caption{(Color online) (a) The Fe-As cluster used in our calculations of the hoppings. 
Green circles are the Fe atoms.
The red circle at the center is an As atom at a distance $c$ below the plane, while the shaded red
circles are the As atoms that are a distance $c$ above the plane. (b) Distances $s$ and $l$ 
for NN Fe-Fe atoms. (c) The distance $d$  along the diagonal
of the Fe-Fe plaquettes. (d) The distance $c$ for As atoms.}
\label{Fig2}
\end{center}
\end{figure}

To guide the discussion, consider a cluster with 4 Fe and 5 As atoms (Fig.~\ref{Fig2}(a)). 
The coordinates of the atoms are needed to calculate hopping amplitudes, 
and they are provided in Table~I,
where $k$, $l$, and $c$ are obtained from the materials structure. The
nearest-neighbor (NN) Fe-Fe distance is $l=2.854$~\AA, \cite{singh} thus $k=l/2=1.427$~\AA. The
distance between Fe and As is $s=2.327$~\AA, \cite{singh} see Fig.~\ref{Fig2}(b).
The next-nearest-neighbor (NNN) Fe-Fe distance along the square diagonal is
$d=\sqrt{2}l=4.037$~\AA (see Fig.~\ref{Fig2}(c)) 
and $r=d/2=2.018$~\AA. According to Fig.~\ref{Fig2}(d),
$c=\sqrt{s^2-r^2}=\sqrt{s^2-l^2/2}=1.158$~\AA.
The director cosines $l$, $m$, and $n$ for
each of the Fe atoms,\cite{slater} with respect to the As located at (0,0,-$c$), are given in
Table~II.

\begin{table}
 \hspace{0.033\textwidth}
 \begin{minipage}{0.2\textwidth}
  \centering

\begin{tabular}{|l|l|l|r|} \hline
Ion & x& y& z\\ \hline
$As_0$ & 0 & 0 & -c \\
$Fe_1$ & k & -k & 0 \\
$Fe_2$ & k & k & 0 \\
$Fe_3$ & -k & k & 0 \\
$Fe_4$ & -k & -k & 0 \\
$As_1$ & l & 0 & c \\
$As_2$ & 0 & l & c \\
$As_3$ & -l & 0 & c \\
$As_4$ & 0 & -l & c \\ \hline
\end{tabular}
  \caption{\textnormal{Coordinates of the atoms in Fig.~\ref{Fig2}(a).}}
  \label{tab:caption1}
 \end{minipage}
 \hspace{0.033\textwidth}
 \begin{minipage}{0.2\textwidth}
  \centering
\begin{tabular}{|l|l|l|r|} \hline
Ion & l& m& n\\ \hline
$Fe_1$ & k/s & -k/s & c/s \\
$Fe_2$ & k/s & k/s & c/s \\
$Fe_3$ & -k/s & k/s & c/s \\
$Fe_4$ & -k/s & -k/s & c/s \\ \hline
\end{tabular}
  \caption{\textnormal{Director cosines of the Fe atoms with respect to As$_0$
in Fig.~\ref{Fig2}(a).}}
  \label{tab:caption2}
 \end{minipage}
\end{table}

\subsection{Overlap integrals between the Fe $d_{\rm xz}$ and $d_{\rm yz}$ orbitals 
and the As $p_{\rm x}$ and $p_{\rm y}$ orbitals}

According to the SK analysis, for the orbitals considered here 
we obtain the following results for the center integrals:

\begin{eqnarray}
E_{x,yz}&=&\sqrt{3}lmn(pd\sigma)-2lmn(pd\pi), \\
E_{y,xz}&=&\sqrt{3}lmn(pd\sigma)-2lmn(pd\pi), \\
E_{x,xz}&=&\sqrt{3}l^2n(pd\sigma)+n(1-2l^2)(pd\pi), \\
E_{y,yz}&=&\sqrt{3}m^2n(pd\sigma)+n(1-2m^2)(pd\pi).
\label{4}
\end{eqnarray}

The corresponding hopping amplitudes are
\begin{eqnarray}
|t_{x,yz}|&=&|t_{y,xz}|=a=\sqrt{3}{k^2c\over{s^3}}(pd\sigma)-2{k^2c\over{s^3}}
(pd\pi),
\label{5}
\end{eqnarray}
\begin{eqnarray}
|t_{x,xz}|=|t_{y,yz}|=b=\sqrt{3}{k^2c\over{s^3}}(pd\sigma)+{c\over{s}}
(1-2{k^2\over{s^2}})(pd\pi). 
\label{6}
\end{eqnarray}
The signs and values of these hoppings for the cluster that we are considering
are in Fig.~3. The values of $l$, $m$, and $n$ shown 
in Table~I are for As$_0$, while some signs will be different for As$_1$, 
As$_2$, As$_3$, and As$_4$.

\begin{figure}[thbp]
\begin{center}
\includegraphics[width=8cm,clip,angle=0]{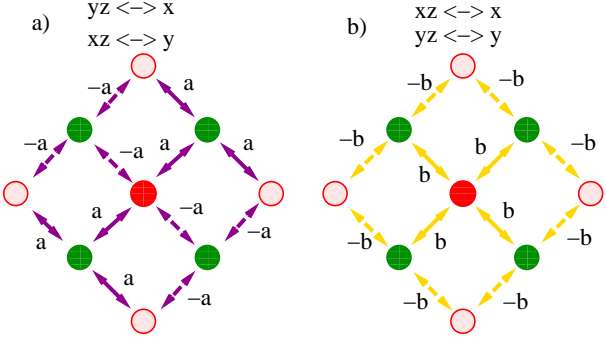}
\vskip 0.3cm
\caption{(Color online) (a) Hoppings between $d_{\rm yz}$ ($d_{\rm xz}$) orbitals in Fe
and $p_{\rm x}$ ($p_{\rm y}$) orbitals in As for the cluster considered in Fig.~\ref{Fig2}(a). (b)
Hoppings between $d_{\rm yz}$ ($d_{\rm xz}$) orbitals in Fe
and $p_{\rm y}$ ($p_{\rm x}$) orbitals in As for the cluster considered in Fig.\ref{Fig2}(a).
Continuous (dashed) lines indicate positive (negative) values.}
\label{Fig3}
\end{center}
\end{figure}

Using the values of $k$, $s$, and $c$ given above, we obtain:
\begin{eqnarray}
a&=&0.324(pd\sigma)-0.374(pd\pi),\\
b&=&0.324(pd\sigma)+0.123(pd\pi).
\label{8}
\end{eqnarray}

Now let us compute the hopping amplitudes for a square lattice made up only of 
Fe atoms. For the NN effective Fe-Fe hopping $t_{nn}$
we will consider the pair of atoms Fe$_1$ and Fe$_2$. For the hopping between the
$d_{\rm xz}$ orbitals, there are two possible paths using the As $p_{\rm y}$ as a bridge. Their
contribution is  given by
(1) $d_{\rm xz}$Fe$_1$-$p_{\rm y}$As$_0$-$d_{\rm xz}$Fe$_2$ 
and (2) $d_{\rm xz}$Fe$_1$-$p_{\rm y}$As$_1$-$d_{\rm xz}$Fe$_2$. From
Fig.~\ref{Fig3}(a), we observe that these paths contribute with $-a^2$ each. Regarding the use
of the $p_{\rm x}$ of As as a bridge, in this case there are also two paths:
(3) $d_{\rm xz}$Fe$_1$-$p_{\rm x}$As$_0$-$d_{\rm xz}$Fe$_2$ and (4) $d_{\rm xz}$Fe$_1$-$p_{\rm x}$As$_1$-$d_{\rm xz}$Fe$_2$. From
Fig.~\ref{Fig3}(b), these paths contribute with $b^2$ each. Reasoning in an analogous manner, four similar
paths are found for the NN hopping between orbitals $d_{\rm yz}$:
(1) $d_{\rm yz}$Fe$_1$-$p_{\rm x}$As$_0$-$d_{\rm yz}$Fe$_2$ and (2) $d_{\rm yz}$Fe$_1$-$p_{\rm x}$As$_1$-$d_{\rm yz}$Fe$_2$, that
from Fig.~\ref{Fig3}(a) they give a contribution $-a^2$ each, and
(3) $d_{\rm yz}$Fe$_1$-$p_{\rm y}$As$_0$-$d_{\rm yz}$Fe$_2$ and
 (4) $d_{\rm yz}$Fe$_1$-$p_{\rm y}$As$_1$-$d_{\rm yz}$Fe$_2$, that from Fig.~\ref{Fig3}(b) 
they give a contribution $b^2$
each. Combining all these results, and 
to second order in perturbation theory,\cite{fulde} 
the Fe-Fe nearest-neighbor hopping
amplitude is given by:
\begin{equation}
t_{nn}^{xz}=t_{nn}^{yz}=(-2a^2+2b^2)/\Delta=2(b^2-a^2)/\Delta,
\label{9}
\end{equation}
\noindent where $\Delta$ is the difference between the on-site energies of the
$d$ and $p$ orbitals.
Notice that by mere geometrical reasons,  
it is not possible to have a nearest-neighbor hopping from
$d_{\rm yz}$ to $d_{\rm xz}$.

For the hopping $t_d$ along the Fe lattice 
plaquette diagonal, namely the NNN Fe-Fe 
hopping,  let us consider the hopping from Fe$_1$ to
Fe$_3$ and from Fe$_2$ to Fe$_4$. It can be easily shown that $d_{\rm xz}$Fe$_1$-$p_{\rm x}$As$_0$-$d_{\rm xz}$Fe$_3$
contributes by an amount  $b^2$ to $t_d^{xz}$, while $d_{\rm xz}$Fe$_1$-$p_{\rm y}$As$_0$-$d_{\rm xz}$Fe$_3$ contributes
$a^2$ to $t_d^{xz}$. The same result is obtained if the hopping from Fe$_2$
to Fe$_4$ is considered. Combining these numbers, then we obtain 
$t_{d}^{xz}=t_{d}^{yz}=(a^2+b^2)/\Delta$.

Along the plaquette diagonal we can also obtain inter-orbital hopping. From Fe$_1$ to
Fe$_3$ the contribution is -$ab$, while from Fe$_2$ to Fe$_4$ it is $ab$. Thus, the
hopping along the $x+y$ and $x-y$ directions are different by a sign from the
inter-orbital hopping. The fact that the plaquette diagonals are equivalent by symmetry
implies that the absolute values of the hoppings must be the same along these diagonals,
but the signs can be different as shown here.
More explicitly, we obtain:
$t_{x+y}^{xz-yz}=ab/\Delta$, and $t_{x-y}^{xz-yz}=-ab/\Delta$.

\subsection{Overlap between $d_{\rm xz}$ and $d_{\rm yz}$ with $p_{\rm z}$}

\begin{figure}[thbp]
\begin{center}
\includegraphics[width=8cm,clip,angle=0]{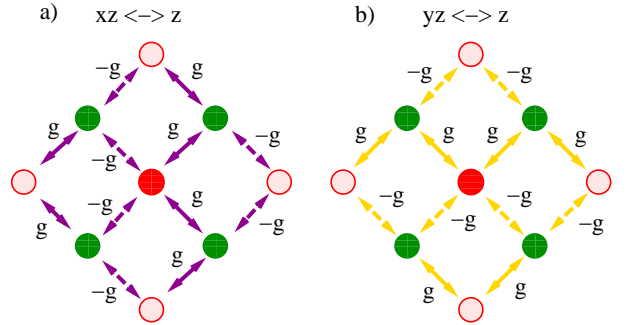}
\vskip 0.3cm
\caption{(Color online) (a) Hoppings between the $d_{\rm yz}$ orbitals in Fe
and the $p_{\rm z}$ orbitals in As, for the cluster considered in Fig.~\ref{Fig2}(a). (b)
Hoppings between the $d_{\rm xz}$ orbitals in Fe
and the $p_{\rm z}$  orbitals in As, for the cluster considered in Fig.~\ref{Fig2}(a).
Continuous (dashed) lines indicate positive (negative) values.}
\label{Fig4}
\end{center}
\end{figure}

The consideration of the $p_{\rm z}$ orbitals adds two new center integrals to the present analysis:
\begin{eqnarray}
E_{z,xz}&=&\sqrt{3}n^2l(pd\sigma)+l(1-2n^2)(pd\pi),\\
E_{z,yz}&=&\sqrt{3}n^2m(pd\sigma)+m(1-2n^2)(pd\pi),
\label{14}
\end{eqnarray}
\noindent which means that a new hopping must be considered
\begin{equation}
|t_{z,xz}|=|t_{z,yz}|=g=\sqrt{3}{kc^2\over{s^3}}(pd\sigma)+
{k\over{s}}(1-2{c^2\over{s^2}})(pd\pi).
\label{15}
\end{equation}
Using the values for $k$, $s$, and $c$ calculated before, 
\begin{equation}
g=0.263(pd\sigma)+0.31(pd\pi).
\label{16}
\end{equation}
\noindent The signs are indicated in Figs.~\ref{Fig4}(a) and (b).

Thus, we obtain an additional contribution to the NN hopping $t_{nn}$
so that $t_{nn}^{xz}=2g^2/\Delta'$ ($-2g^2/\Delta'$) along the $y$ ($x$) axis. Reciprocally, 
$t_{nn}^{yz}=2g^2/\Delta'$ ($-2g^2/\Delta'$) along the $x$ ($y$) axis.
Along the diagonal,
$t_d=-g^2/\Delta'$ for both orbitals is obtained. Note also that $p_z$ generates
an inter-orbital diagonal hopping given by $-g^2/\Delta'$ ($g^2/\Delta'$)
along the $x+y$ $(x-y)$ directions. $\Delta'$ is the difference between
the on-site energies of the $d$ and $p_z$ orbitals. From
Ref.~\onlinecite{cao}, the gaps are $\Delta=1.25$~eV and $\Delta'=5$~eV, but
other values for these gaps are also considered below.

\subsection{Direct Fe-Fe hopping}

Since the distance between Fe atoms is $l=2.854$~\AA, comparable to the
Fe-As distance, the contributions to the electron hoppings
coming from the direct overlap between the $d$ orbitals of the Fe atoms should also be
considered.
Following SK,\cite{slater}
$E_{xz,xz}=3l^2n^2(dd\sigma)+(l^2+n^2-4l^2n^2)(dd\pi)+
(m^2+l^2n^2)dd\delta$,
$E_{yz,yz}=3m^2n^2(dd\sigma)+(m^2+n^2-4m^2n^2)(dd\pi)+
(l^2+m^2n^2)dd\delta,$ and
$E_{xz,yz}=3lmn^2(dd\sigma)+lm(1-4n^2)[(dd\pi)-(dd\delta)]$.
Notice that all the Fe atoms have $n=0$, and $l=\pm1$, $m=0$ ($l=0$, $m=\pm1$) if
they are neighbors along the $x$ ($y$) direction. 
Thus, the inter-orbital hopping vanishes, and
we obtain $t_{xz,xz}=-dd\pi$ ($t_{yz,yz}=-dd\pi$) along the direction $x$ ($y$), 
and $dd\delta$ along $y$ ($x$). These same expressions can be used to obtain 
the diagonal Fe-Fe hopping parameters. We find that 
$t^d_{\alpha,\alpha}=-(dd\pi'+dd\delta')/2$, where $\alpha=xz$ or $yz$, while
$t^d_{xz,yz}=\pm(dd\pi'-dd\delta')/2$ with the minus (plus) sign for the 
$\hat x+\hat y$ ($\hat x-\hat y$) direction and the prime indicates second 
nearest-neighbors overlap integrals.

\section{Effective two-orbital tight-binding model}

\subsection{Hopping Term}

Considering the results of the previous section, the
kinetic-energy term of the 
effective tight-binding Hamiltonian involving the $d_{\rm xz}$ and $d_{\rm yz}$
orbitals, defined on the square lattice formed only by the Fe atoms, is given by:
$$
H_{\rm TB}=-t_1\sum_{{\bf i},\sigma}(d^{\dagger}_{{\bf i},x,\sigma}
d_{{\bf i}+\hat y,x,\sigma}+d^{\dagger}_{{\bf i},y,\sigma}
d_{{\bf i}+\hat x,y,\sigma}+h.c.)
$$
$$
-t_2\sum_{{\bf i},\sigma}(d^{\dagger}_{{\bf i},x,\sigma}
d_{{\bf i}+\hat x,x,\sigma}+d^{\dagger}_{{\bf i},y,\sigma}
d_{{\bf i}+\hat y,y,\sigma}+h.c.)
$$
$$
-t_3\sum_{{\bf i},\hat\mu,\hat\nu,\sigma}(d^{\dagger}_{{\bf i},x,\sigma}
d_{{\bf i}+\hat\mu+\hat\nu,x,\sigma}+d^{\dagger}_{{\bf i},y,\sigma}
d_{{\bf i}+\hat\mu+\hat\nu,y,\sigma}+h.c.)
$$
$$
+t_4\sum_{{\bf i},\sigma}(d^{\dagger}_{{\bf i},x,\sigma}
d_{{\bf i}+\hat x+\hat y,y,\sigma}+d^{\dagger}_{{\bf i},y,\sigma}
d_{{\bf i}+\hat x+\hat y,x,\sigma}+h.c.)
$$
$$
-t_4\sum_{{\bf i},\sigma}(d^{\dagger}_{{\bf i},x,\sigma}
d_{{\bf i}+\hat x-\hat y,y,\sigma}+d^{\dagger}_{{\bf i},y,\sigma}
d_{{\bf i}+\hat x-\hat y,x,\sigma}+h.c.)
$$
\begin{equation}
-\mu\sum_{\bf i}(n^x_{\bf i}+n^y_{\bf i}).
\label{20}
\end{equation}


In this Hamiltonian, the operator $d^{\dagger}_{{\bf i},\alpha,\sigma}$ creates
an electron with spin $z$-axis projection $\sigma$, 
orbital $\alpha$, and on the site ${\bf i}$ of a square lattice. The
chemical potential is given by $\mu$ and $n^{\alpha}_{\bf i}$ are number
operators. The index $\hat\mu=\hat x$ or $\hat y$ 
is a unit vector linking nearest-neighbor sites. The
hoppings,
within the SK approach, are given by:
\begin{eqnarray}
t_1&=&-2[(b^2-a^2)/\Delta+g^2/\Delta']-dd\delta, \nonumber \\
t_2&=&-2[(b^2-a^2)/\Delta-g^2/\Delta']-dd\pi, \nonumber \\
t_3&=&-[(a^2+b^2\Delta-g^2/\Delta')]-(dd\pi'+dd\delta')/2, \nonumber  \\
t_4&=&-(ab/\Delta-g^2/\Delta')-(dd\pi'-dd\delta')/2. 
\label{24}
\end{eqnarray}

\noindent The explicit expressions for these hopping amplitudes in
terms of the overlap integrals using the parameters for FeAs 
can be easily found and they will not be provided here.
The two orbital model proposed by Raghu {\it et al.}\cite{scalapino} has the 
same form as
the one presented above but the hoppings
are obtained by fitting band structures.\cite{fang2} 

It is interesting to notice that if only the direct 
overlap between the $d$ orbitals is considered, i.e. ignoring the indirect 
hopping through the $p$ As orbitals, the form of Eq.~(\ref{20}) does
not change. Thus, the form of $H_{\rm TB}$ arises from the symmetry properties of
the $d_{\rm xz}$ and $d_{\rm yz}$ orbitals rather than from the location of the As 
ions. However, the indirect Fe-Fe hopping through the As atoms plays a key 
role in providing the relatively large value of the diagonal hopping $t_3$ vs. the 
NN hoppings which, as discussed below, 
stabilizes the magnetic stripe order. For example, if we only consider the 
direct hopping then $t_3/t_2\approx dd\pi'/2dd\pi$, where  
$dd\delta'\approx 0$ was assumed.\cite{Harrison} Since $dd\pi'\ll dd\pi$, 
then $|t_3|\ll|t_2|$. 
However, if we consider the indirect hopping then $|t_3|\geq|t_2|/2$.

\begin{figure}[thbp]
\begin{center}
\includegraphics[width=8cm,clip,angle=0]{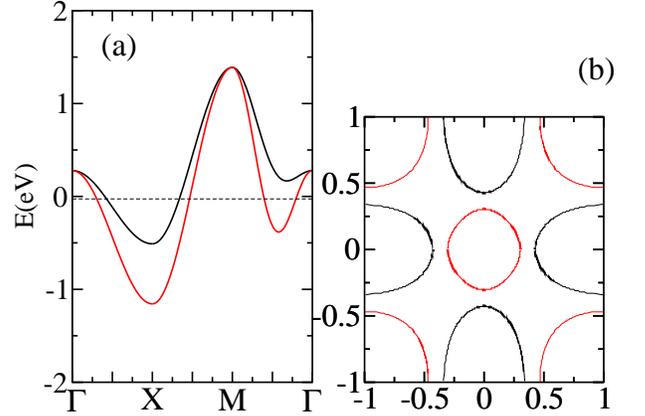}
\vskip 0.3cm
\caption{(Color online) (a) Energy vs. momentum for the non-interacting 
tight-binding Hamiltonian
in Eq.~(\ref{20}) using $t_1=0.058$~eV, $t_2=0.22$~eV, $t_3=-0.21$~eV, and
$t_4=-0.08$~eV. These hopping amplitudes are
obtained from the Slater-Koster formulas using $(pd\sigma)=1$~eV and $(pd\pi)=-0.2$~eV, 
supplemented by $\Delta=\Delta'=1$~eV for simplicity, and $\mu=-0.03$~eV, which
corresponds to half-filled orbitals. Results are plotted along the path $(0,0)-(\pi,0)-
(\pi,\pi)-(0,0)$. (b) Fermi surface for the half-filled system.}
\label{set1}
\end{center}
\end{figure}

\begin{figure}[thbp]
\begin{center}
\includegraphics[width=8cm,clip,angle=0]{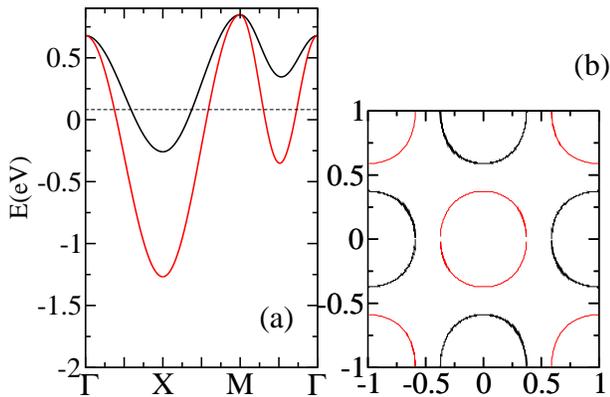}
\vskip 0.3cm
\caption{(Color online) (a) Energy vs. momentum for the non-interacting 
tight-binding Hamiltonian
Eq.~(\ref{20}) using $t_1=-0.1051$~eV, $t_2=0.1472$~eV, $t_3=-0.1909$~eV, and
$t_4=-0.0874$~eV,
obtained using the parameters $(pd\sigma)=1$~eV, $(pd\pi)=-0.2$~eV, $dd\pi=0.2$~eV,
$dd\delta=-0.02$~eV, $\Delta=1.25$~eV, and $\Delta'=5$~eV. The chemical potential is 
$\mu=0.081$~eV, which
corresponds to half-filled orbitals. Results are plotted along the path $(0,0)$-$(\pi,0)$-
$(\pi,\pi)$-$(0,0)$. (b) Fermi surface for the half-filled system.}
\label{set2}
\end{center}
\end{figure}

To analyze the influence of the several parameters, let us 
consider two special cases.
Setting $(pd\sigma)=1.0$~eV, $pd\pi=-0.2$~eV, $\Delta=1.0$~eV, and $\Delta'$=1~eV 
in Eqs.~(15-18), and
neglecting the direct Fe-Fe coupling i.e. using
$dd\pi=dd\delta=0$, we obtain: $t_1=0.058$~eV, $t_2=0.22$~eV, $t_3=-0.21$~eV, and
$t_4=-0.08$~eV. With these values, the band structure, shown in Fig.~\ref{set1}, is
qualitatively similar to the band-structure calculations,
although
the pockets are larger in size.
%
%
Another example can be obtained by using 
the calculated values of the energy gaps, which are $\Delta=1.25$~eV and
$\Delta'=5$~eV.\cite{cao} In Fig.~\ref{set2}, the band structure is shown for
$(pd\sigma)=1$~eV, $pd\pi=-0.2$~eV, $\Delta=1.25$~eV, $\Delta'=5$~eV, 
$dd\pi=0.2$~eV, and $dd\delta=-0.02$~eV, i.e. including the direct 
Fe-Fe hopping. Now
the hole pocket at $\Gamma$ is larger than in the previous case, but the 
overall shape remains similar.

\begin{figure}[thbp]
\begin{center}
\includegraphics[width=8cm,clip,angle=0]{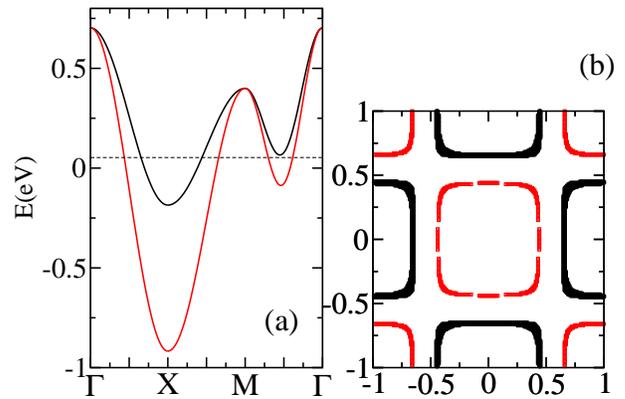}
\vskip 0.3cm
\caption{(Color online) (a) Energy vs. momentum for the non-interacting 
tight-binding Hamiltonian
in Eq.~(\ref{20}) using $t_1=-0.129$~eV, $t_2=0.05$~eV, $t_3=-0.137$~eV, and
$t_4=-0.019$~eV obtained for $(pd\sigma)=-0.41$~eV, $(pd\pi)=0.19$~eV, 
$dd\pi=0.18$~eV, $dd\delta=0$, $\Delta=1.25$~eV, and $\Delta'=5$~eV.
The chemical potential is $\mu=0.053$~eV, which
corresponds to half-filled orbitals. Results are plotted along the path $(0,0)$-$(\pi,0)$-
$(\pi,\pi)$-$(0,0)$. (b) Fermi surface for the half-filled system.}
\label{set7}
\end{center}
\end{figure}

Notice that the overlap integrals can also be estimated using 
tabulated values and the distances between the atoms.\cite{Harrison} 
The band structure and Fermi surface obtained using these values are shown in Fig.~\ref{set7}.

\begin{figure}[thbp]
\begin{center}
\includegraphics[width=8cm,clip,angle=0]{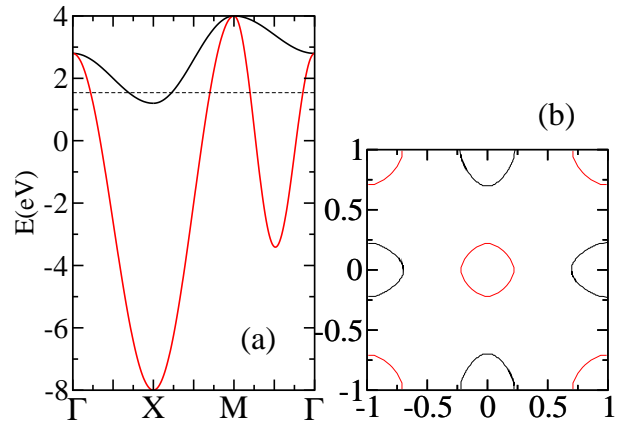}
\vskip 0.3cm
\caption{(Color online)
(a) Energy vs. momentum for the non-interacting 
tight-binding Hamiltonian
Eq.~(\ref{20}) using the hopping amplitudes obtained from
fits of band-structure calculations:\cite{scalapino} $t_1=-1.0$, $t_2=1.3$,
and $t_3=t_4=-0.85$ (all in eV units).
Results are plotted along the path $(0,0)$-$(\pi,0)$-
$(\pi,\pi)$-$(0,0)$. 
(b) Fermi surface for the half-filled system.
}
\label{set3}
\end{center}
\end{figure}

To complete the analysis, let us discuss now the results obtained using the set of hoppings 
that fit band-structure calculations.\cite{scalapino} 
The dispersion
and Fermi surface are in Fig.~\ref{set3}.
By construction, the agreement with the band-structure Fermi surface is better than
in the other cases, the main difference being the size of the pockets. Nevertheless,
it appears that in a broad range of hoppings and couplings, the qualitative topology of
the Fermi surfaces remains the same, and this is probably the reason why the main
magnetic and pairing properties are also similar among the many sets, as shown explicitly below.

\subsection{Interactions}

In this section, the Coulombic interaction terms are added to the tight-binding
Hamiltonian Eq.~(\ref{20}) to form the full two-orbital model. These Coulombic terms are:\cite{ours}
$$
H_{\rm int}=
U\sum_{{\bf i},\alpha}n_{{\bf i},\alpha,\uparrow}n_{{\bf i},
\alpha,\downarrow}
+(U'-J/2)\sum_{{\bf i}}n_{{\bf i},x}n_{{\bf i},y}
$$
\begin{equation}
-2J\sum_{\bf i}{\bf S}_{{\bf i},x}\cdot{\bf S}_{{\bf i},y}
+J\sum_{{\bf i}}(d^{\dagger}_{{\bf i},x,\uparrow}
d^{\dagger}_{{\bf i},x,\downarrow}d_{{\bf i},y,\downarrow}
d_{{\bf i},y,\uparrow+h.c.)},
\label{29}
\end{equation}

\noindent where $\alpha=x,y$ denotes the orbital, ${\bf S}_{{\bf i},\alpha}$
($n_{{\bf i},\alpha}$) is the spin (electronic density) in orbital $\alpha$ at site
${\bf i}$, and we have used the relation $U'=U-2J$, from 
rotational invariance.\cite{RMP01}

\subsection{Pairing}

Diagonalizing exactly the full two-orbital Hamiltonian 
on a $\sqrt{8}\times \sqrt{8}$ cluster with periodic boundary
conditions, it was observed in previous investigations that 
in regions of parameter space the ground state with two extra electrons above
half-filling is a spin triplet.\cite{ours} In this case, the relevant pairing operator is given by:
\begin{equation}
\Delta^{\dagger}({\bf i})_{\sigma}=\sum_{\mu}(d^{\dagger}_{{\bf i},x,\sigma}
d^{\dagger}_{{\bf i}+\mu,y,\sigma}-
d^{\dagger}_{{\bf i},y,\sigma}
d^{\dagger}_{{\bf i}+\mu,x,\sigma}),
\label{30}
\end{equation}

\noindent where $\sigma=\uparrow$ or $\downarrow$ denotes the
spin projection 1 or -1, respectively, while the 0 projection operator is:

\begin{eqnarray}
\Delta^{\dagger}({\bf i})_0=\sum_{\mu}(d^{\dagger}_{{\bf i},x,\uparrow}
d^{\dagger}_{{\bf i}+\mu,y,\downarrow}&+&
d^{\dagger}_{{\bf i},x,\downarrow}
d^{\dagger}_{{\bf i}+\mu,y,\uparrow} - \nonumber \\
d^{\dagger}_{{\bf i},y,\uparrow}
d^{\dagger}_{{\bf i}+\mu,x,\downarrow}
&-&d^{\dagger}_{{\bf i},y,\downarrow}
d^{\dagger}_{{\bf i}+\mu,x,\uparrow}),
\label{32}
\end{eqnarray}

\noindent or, in momentum space,
\begin{equation}
 \Delta^{\dagger}({\bf k})_{\sigma}=(\cos k_x+\cos k_y)
(d^\dagger_{{\bf k},x,\sigma} d^\dagger_{{\bf -k},y,\sigma}-
 d^\dagger_{{\bf k},y,\sigma} d^\dagger_{{\bf -k},x,\sigma}),
\label{33}
\end{equation}


\begin{eqnarray}
&& \Delta^{\dagger}({\bf k})_0=(\cos k_x+\cos k_y)
(d^\dagger_{{\bf k},x,\uparrow} d^\dagger_{{\bf -k},y,\downarrow} + \nonumber \\
%
&&d^\dagger_{{\bf k},x,\downarrow} d^\dagger_{{\bf -k},y,\uparrow}- 
 d^\dagger_{{\bf k},y,\uparrow} d^\dagger_{{\bf -k},x,\downarrow}-
 d^\dagger_{{\bf k},y,\downarrow} d^\dagger_{{\bf -k},x,\uparrow}).
\label{35}
\end{eqnarray}

\noindent
This operator is invariant under the $A_{\rm 2g}$ irreducible representation
of the group $D_{\rm 4h}$, it is odd under orbital
exchange, and it is a spin triplet.

However, in our previous effort we have also identified 
regions of parameter space where the state with two extra
electrons is a spin {\it singlet}, which appears to be compatible with
the results of experiments that favor singlet states over triplets.\cite{Grafe,matano,kawabata} 
The dominant pairing operator 
for the singlet is given by:
\begin{eqnarray}
\Delta^{\dagger}({\bf i})=\sum_{\alpha}d^{\dagger}_{{\bf i},\alpha,\uparrow}
(d^{\dagger}_{{\bf i}+\hat x,-\alpha,\downarrow}&+&
d^{\dagger}_{{\bf i}+\hat y,-\alpha,\downarrow}+ \nonumber \\
d^{\dagger}_{{\bf i}-\hat x,-\alpha,\downarrow}&+&
d^{\dagger}_{{\bf i}-\hat y,-\alpha,\downarrow}),
\label{36}
\end{eqnarray}
\noindent that in momentum space becomes
\begin{equation}
\Delta^{\dagger}({\bf k})=\sum_{\alpha}(\cos k_x+\cos k_y)
d^{\dagger}_{{\bf k},\alpha,\uparrow}d^{\dagger}_{{\bf -k},-\alpha,\downarrow}.
\label{37}
\end{equation}

This operator transforms as the $B_{\rm 2g}$ irreducible representation of the
$D_{\rm 4h}$ point group, it is even under orbital exchange, and it is a spin singlet.
As explained in the introduction, the experimental results favoring spin singlet
pairing lead us to focus our effort on this spin-singlet operator 
in the following sections.

\section{Exact Diagonalization results}

\subsection{Method}

In  this section, the Lanczos or Exact Diagonalization (ED) method will be used 
to obtain the ground 
state of the two-orbital model, both at half filling and also for a system with 
two electrons more than half filling. Due to the exponential growth of the Hilbert
space with increasing cluster sizes, here our effort must be 
restricted to a tilted $\sqrt{8}$$\times$$\sqrt{8}$
cluster.\cite{ours} Using translational 
invariance, the Hilbert space can be reduced to $21,081,060$ states at 
half filling and $16,359,200$ for two electrons away from half filling. 
Taking into account the additional symmetries of spin inversion as well 
as rotations, the dimension of the Hilbert space becomes $\approx 2,600,000$. 
The employed Lanczos scheme is standard and requires up to 11~GB of memory
when only translational invariance is used. 
Note that the two-orbital 8-sites cluster has a similar 
Hilbert-space size as a 16 sites one-band Hubbard lattice, and they are similarly
computationally demanding. 
The focus of our effort is on ground states for a fixed set of quantum
numbers corresponding to the symmetries that were implemented.
We use both the hoppings from the SK approach 
and also the hoppings that fit band-structure calculations, and 
find qualitatively consistent results for both sets.


\begin{figure}
\subfiguretopcaptrue
\subfigure[$\quad pd\pi/pd\sigma=-0.5$]{\includegraphics[width=0.23\textwidth]{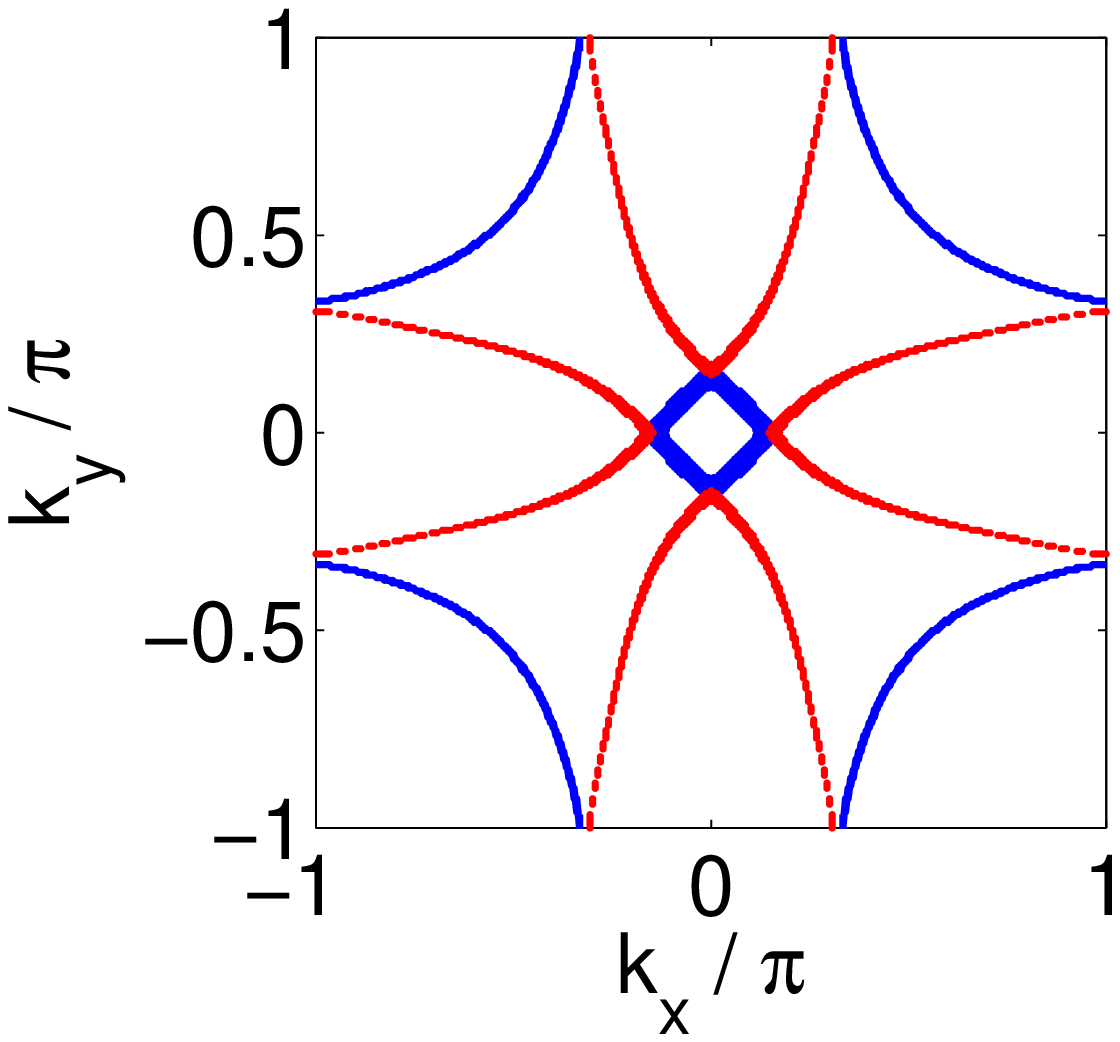}\label{fig:fs_pdp-05}}   
\subfigure[$\quad pd\pi/pd\sigma=-0.2$]{\includegraphics[width=0.23\textwidth]{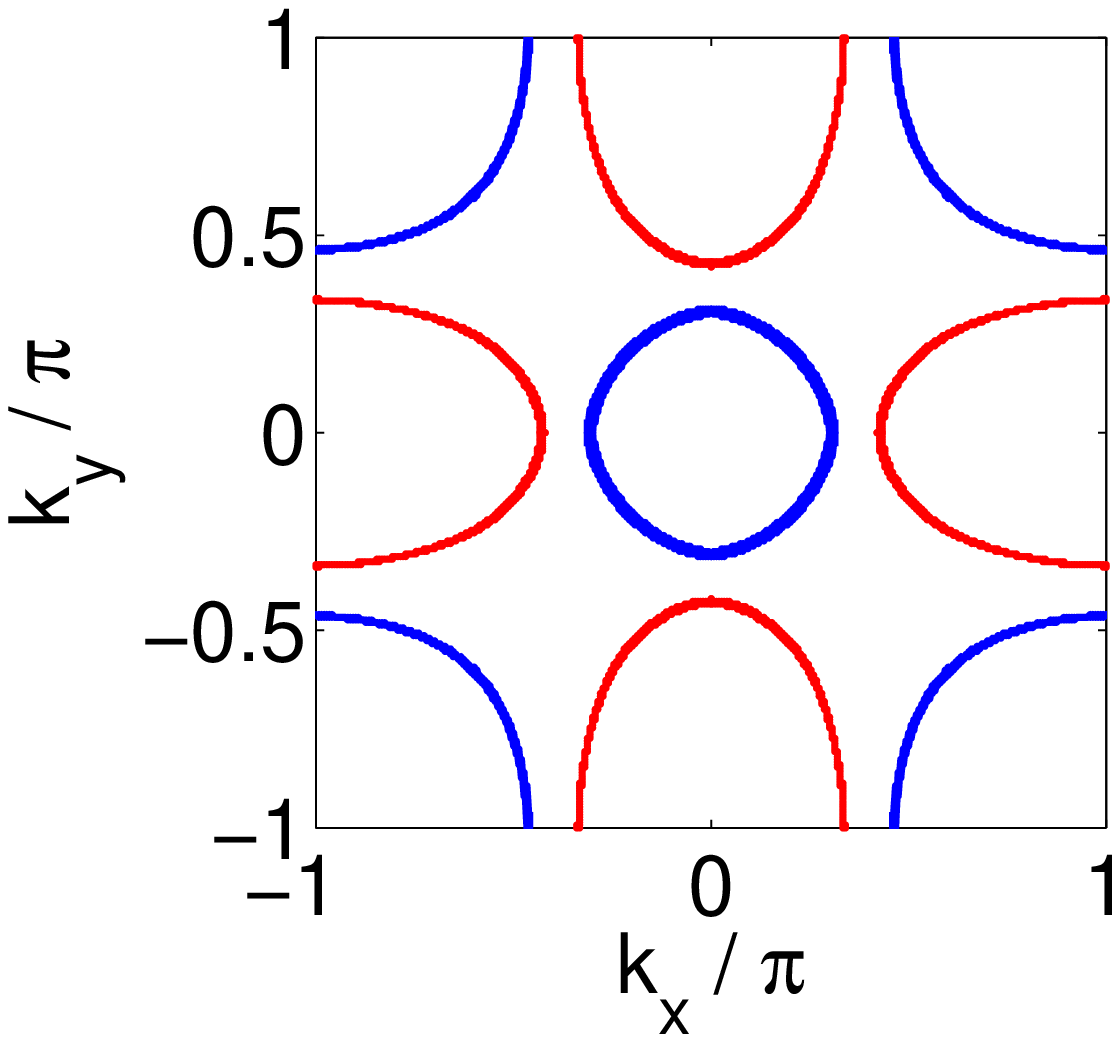}\label{fig:fs_pdp-02}}\\[-0.5em]
\subfigure[$\quad pd\pi/pd\sigma=-0.1$]{\includegraphics[width=0.23\textwidth]{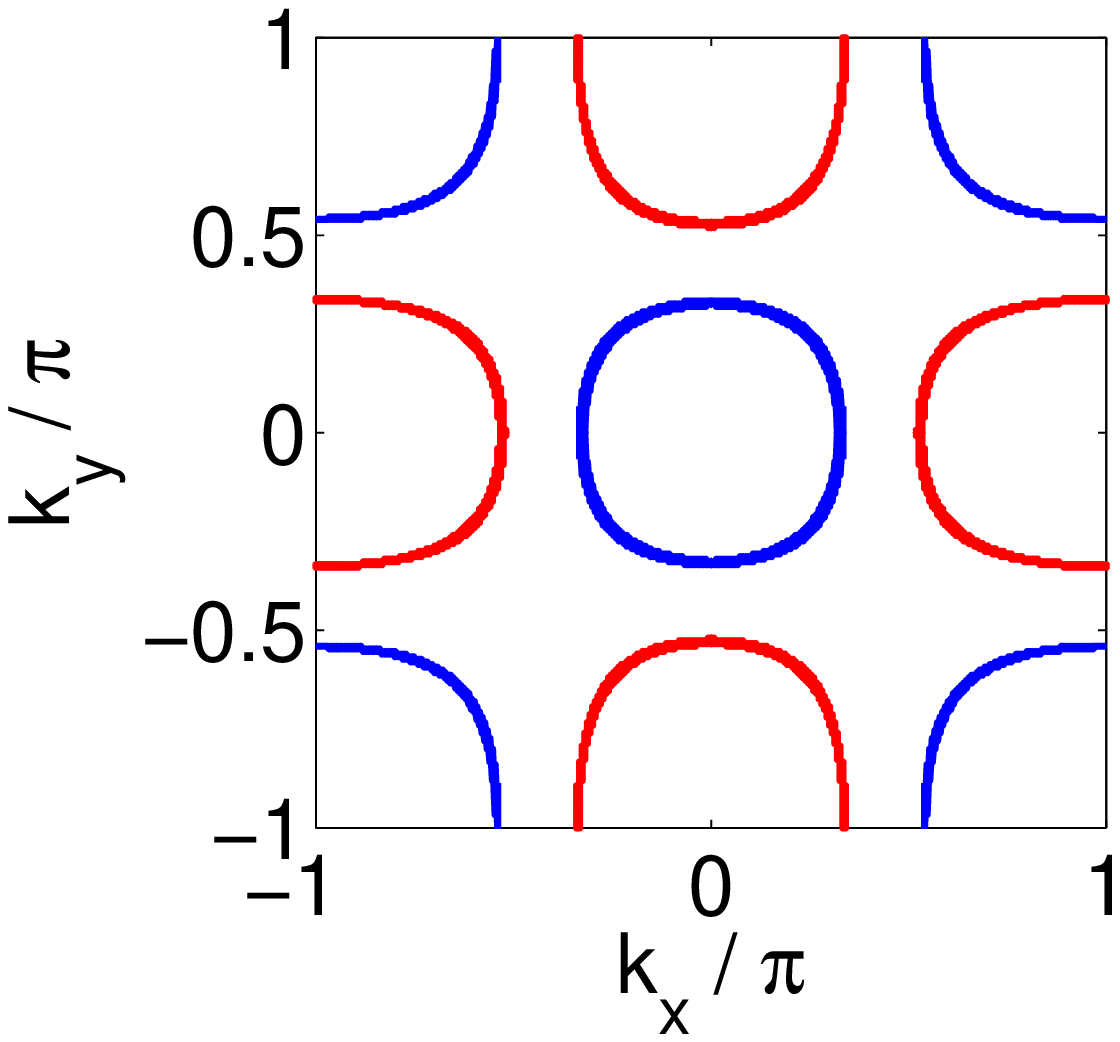}\label{fig:fs_pdp-01}}   
\subfigure[$\quad pd\pi=0$]{\includegraphics[width=0.23\textwidth]{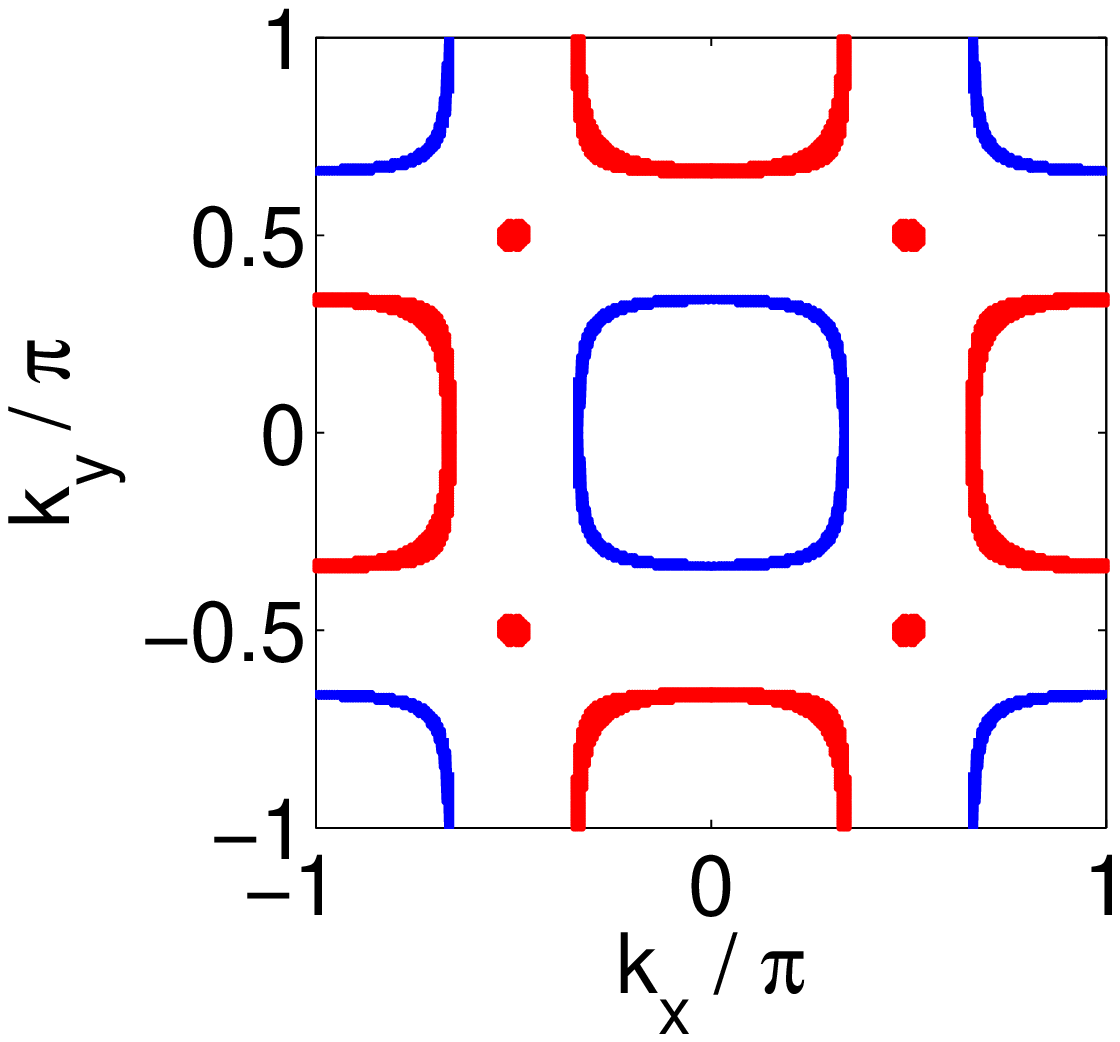}\label{fig:fs_pdp0}}\\[-0.5em]
\subfigure[$\quad pd\pi/pd\sigma=0.1$]{\includegraphics[width=0.23\textwidth]{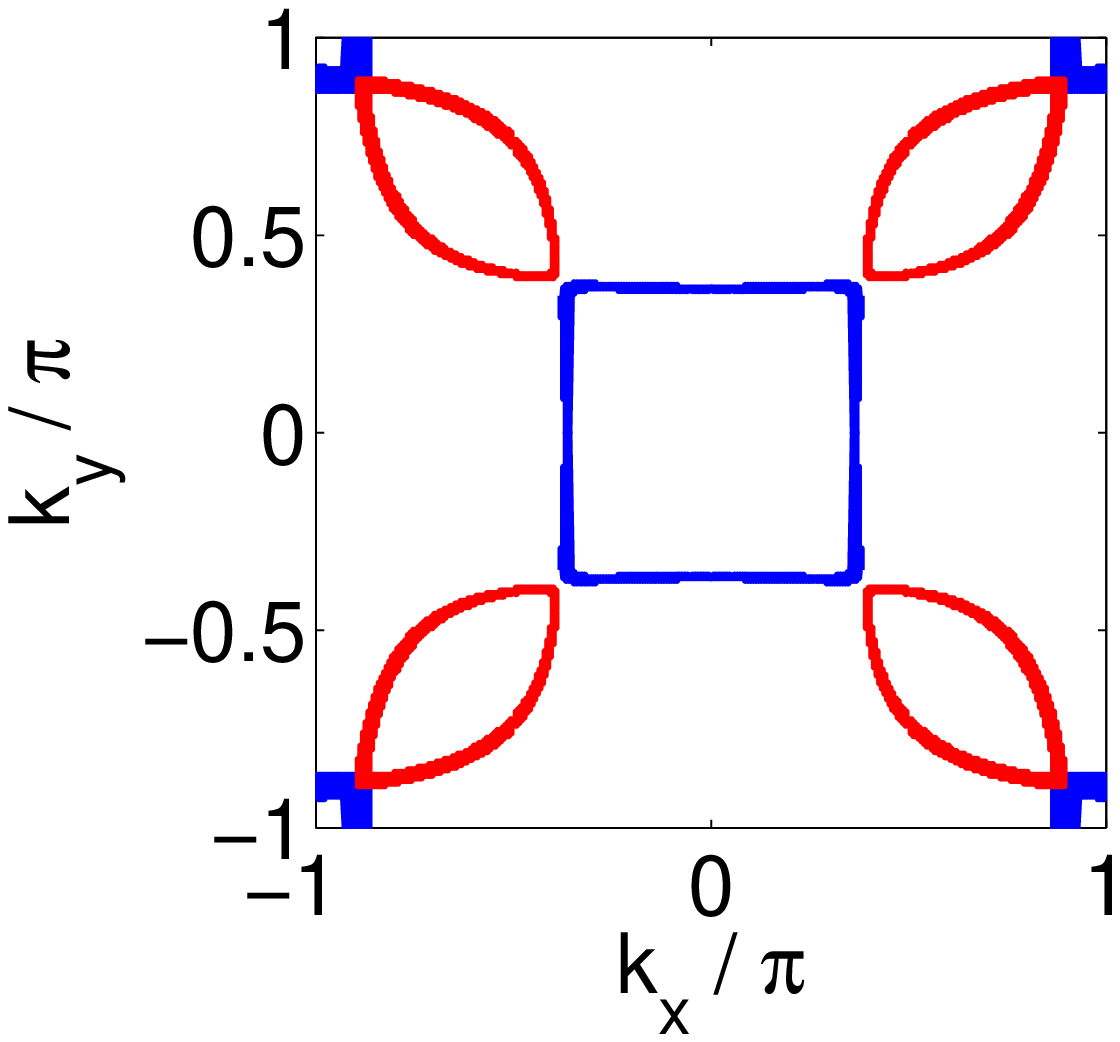}\label{fig:fs_pdp01}}   
\subfigure[$\quad pd\pi/pd\sigma=0.5$]{\includegraphics[width=0.23\textwidth]{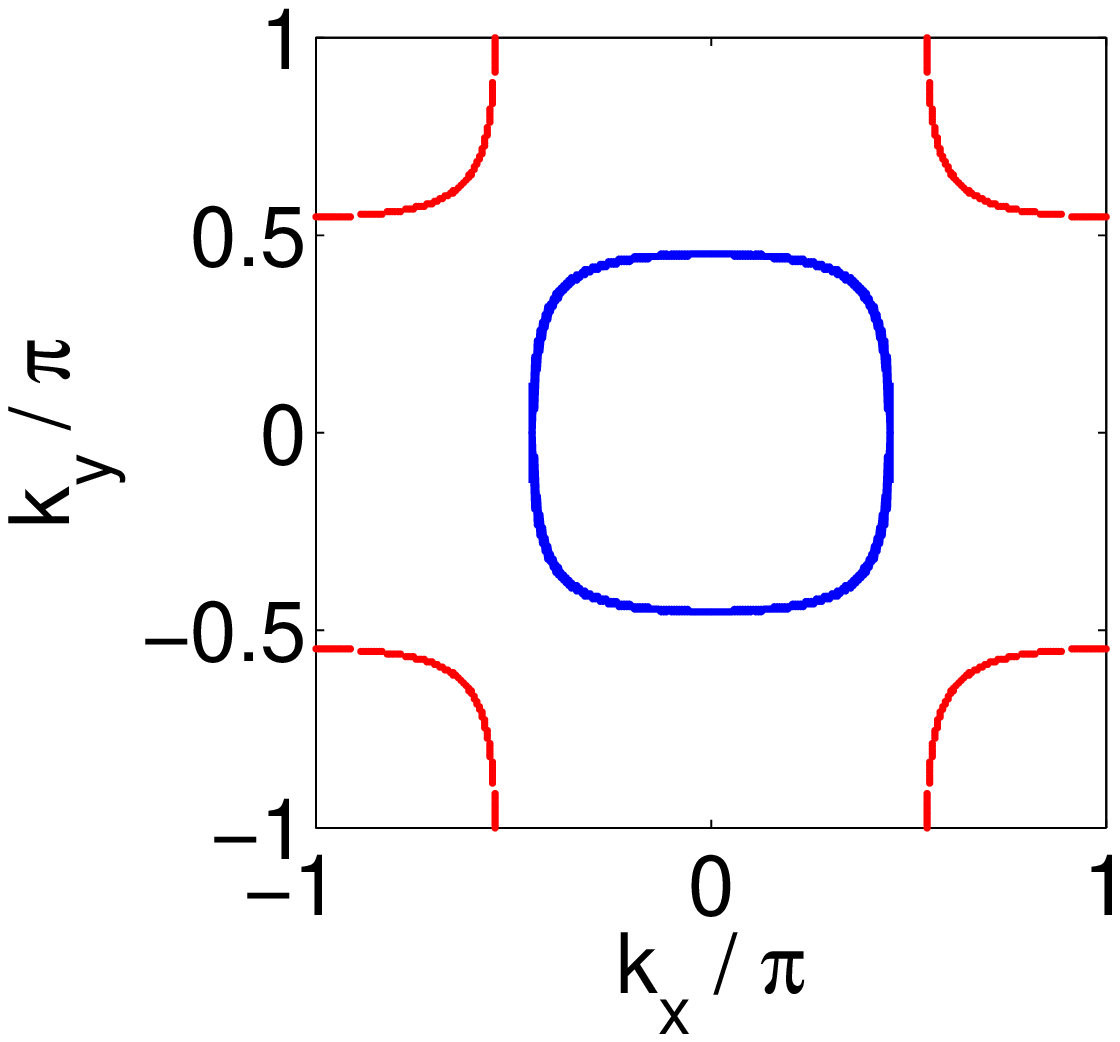}\label{fig:fs_pdp05}}\\
\caption{(Color online) Non-interacting ($U=J=0$) 
Fermi surface   in the unfolded
  Brillouin zone, at the 
$pd\pi/pd\sigma$s indicated.
The
realistic regime is  $pd\pi/pd\sigma < 0$ since 
  hole pockets around $(0,0)$  and $(\pi,\pi)$, and electron
  pockets around $(0,\pi)$ and  $(0,\pi)$, are observed. 
\label{fig:fs_pdp} }    
\end{figure}

\subsection{Results using Slater-Koster derived hoppings}

\subsubsection{Fermi surfaces, spin order, and spin of the pairs}

\begin{figure}
\subfigure{\includegraphics[width=0.23\textwidth]{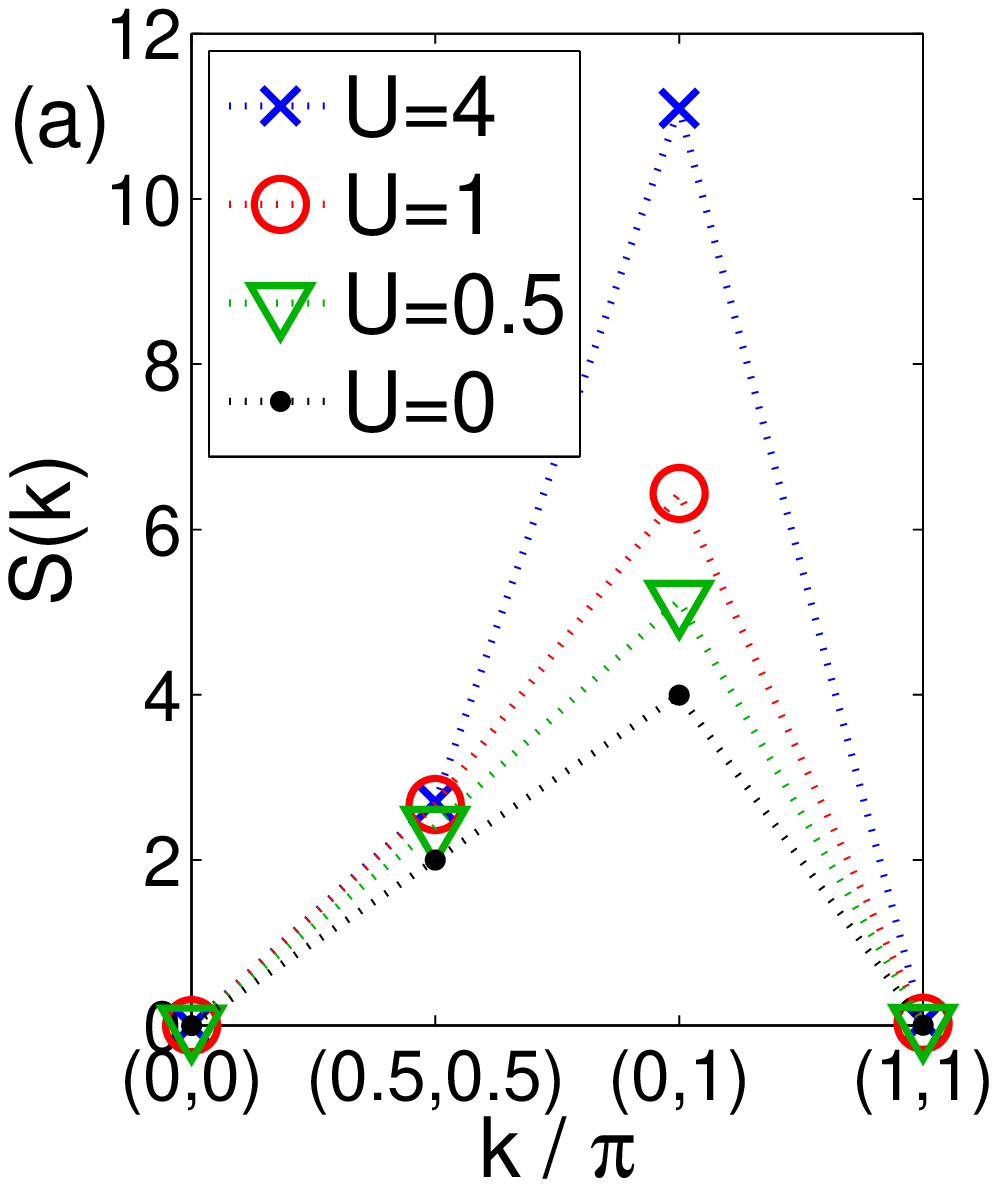}\label{fig:sk_U}}
\subfigure{\includegraphics[width=0.23\textwidth]{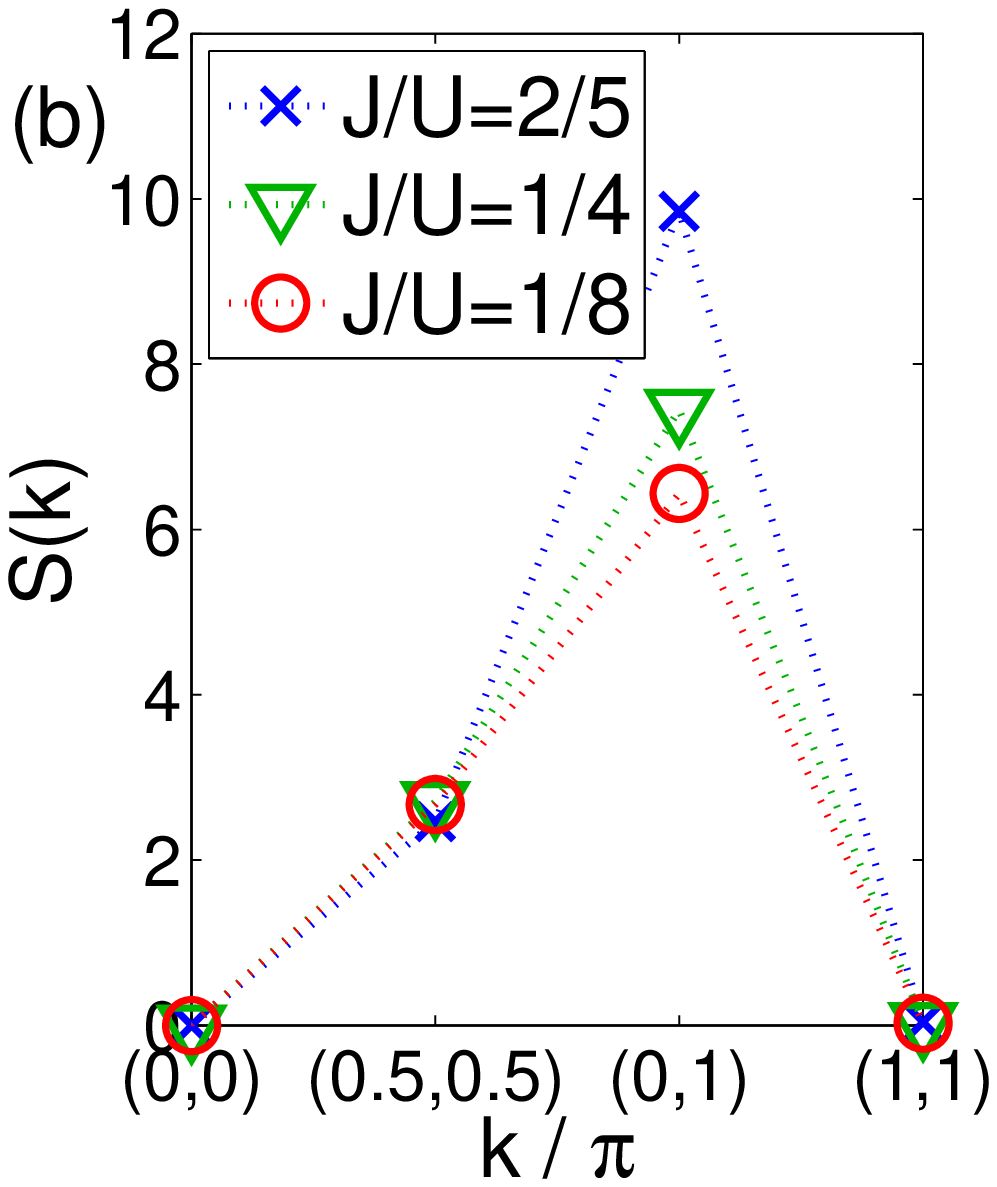}\label{fig:sk_J}}
\caption{(Color online) Spin structure factor $S(k)$ at half filling obtained with ED 
on  a $\sqrt{8}\times\sqrt{8}$ cluster for
  $pd\pi/pd\sigma = -0.2$. (a) Results corresponding to several $U$'s and $J/U=1/8$. 
(b) Results varying $J$, at fixed $U=1$.\label{fig:sk_U_J_pdp-0.2} }  
\end{figure}

In the SK approach, the parameter $pd\sigma$ is here kept 
fixed equal to 1, providing the scale, and 
the free parameter $pd\pi$ is varied. To constrain the values of $pd\pi$, let us return to the
tight-binding Hamiltonian. Figure~\ref{fig:fs_pdp} shows how the Fermi
surface evolves by changing $pd\pi$. These figures are in the
unfolded Brillouin zone, i.e., for one Fe atom per unit cell. For
a negative ratio $pd\pi/pd\sigma$, 
hole pockets around momenta $(0,0)$
and $(\pi,\pi)$ and electron pockets around $(0,\pi)/(\pi,0)$ are found. However, for
a vanishing ratio
$pd\pi=0$, additional electron pockets appear at $(\pi/2, \pi/2)$, while
the pockets around $(0,\pi)$ and $(\pi,0)$ 
disappear fast by further increasing $pd\pi$ to positive
values. As discussed before,\cite{ours} the robust  
NNN
hopping $t_3$ at negative $pd\pi/pd\sigma$  induces tendencies toward a $(0,\pi) / 
(\pi, 0)$ magnetic ordering at half filling, as shown by the spin structure
factor in Fig.~\ref{fig:sk_U}. This is in good agreement with neutron
scattering experiments. Thus, it is clear that 
the realistic regime corresponds to negative $pd\pi$, and an opposite
sign of $pd\pi$ and $pd\sigma$ is also what would be expected from
the tabulated values.\cite{Harrison} 

\begin{figure}
\subfigure{\includegraphics[width=0.23\textwidth]{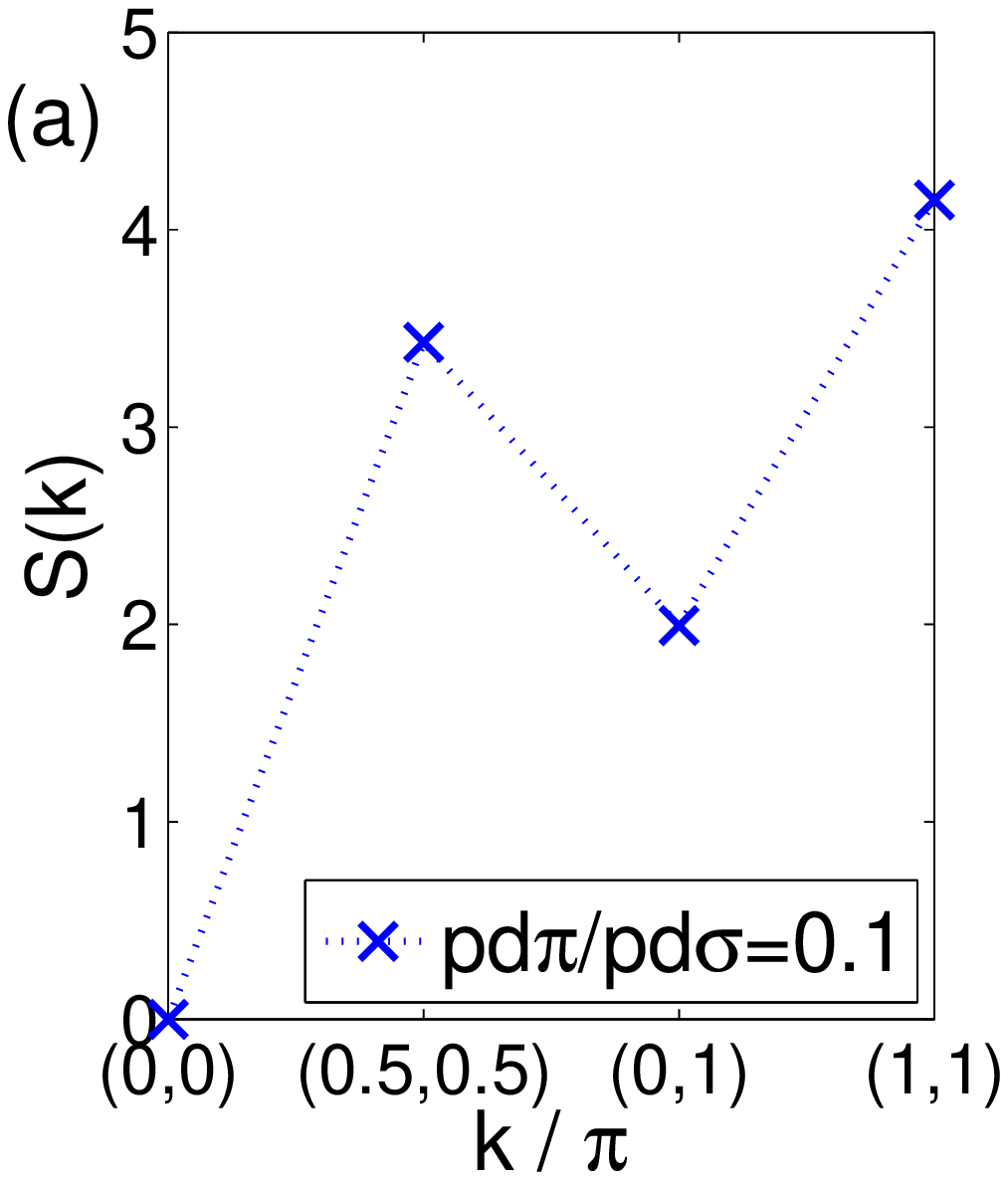}\label{fig:sk_pdpi0.1}}
\subfigure{\includegraphics[width=0.23\textwidth]{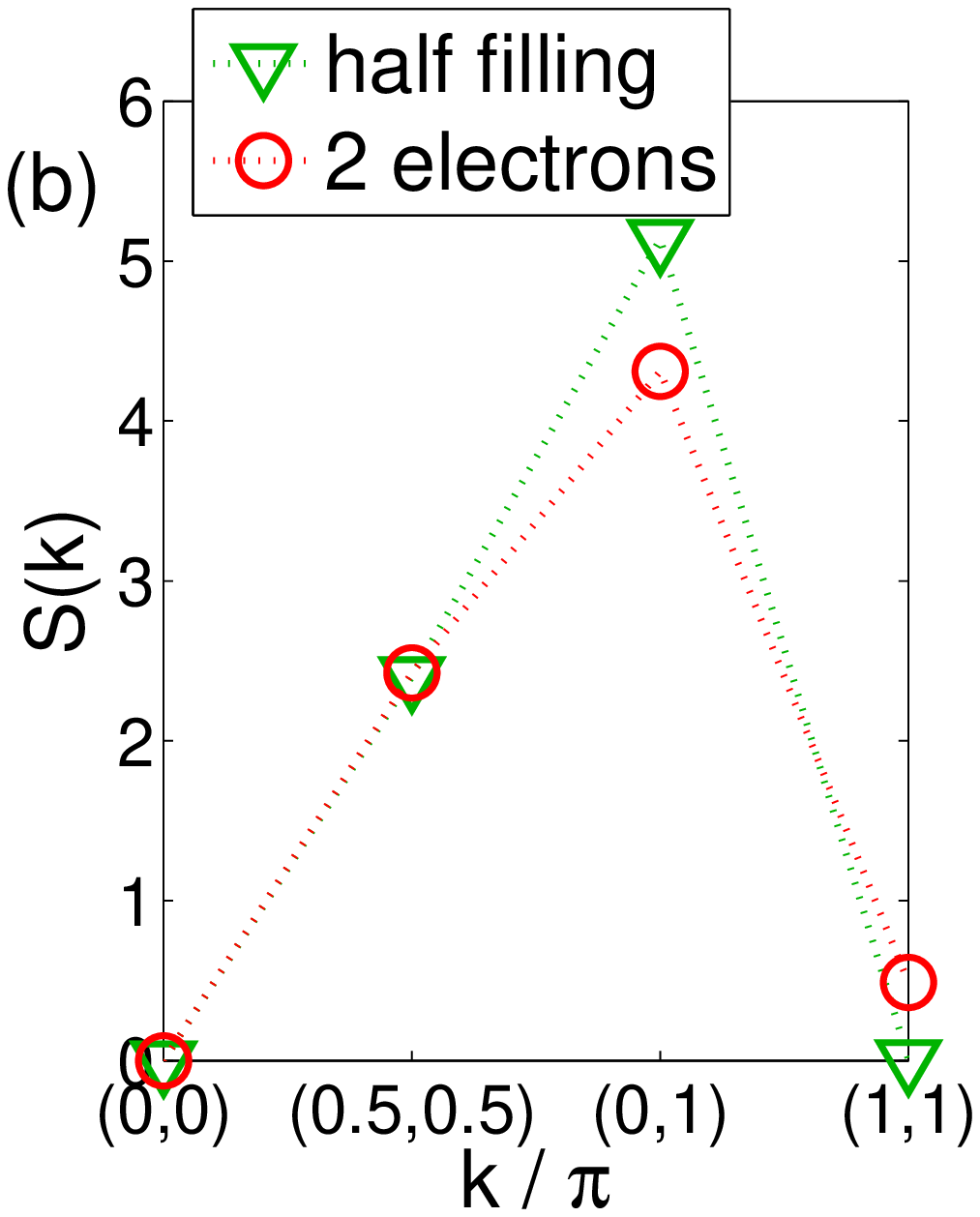}\label{fig:sk_J_doped}}
\caption{(Color online) Spin structure factor $S(k)$ for (a) half
  filling and $pd\pi/pd\sigma=0.1$, $U=0.5$, $J=U/8$, 
and (b) $pd\pi/pd\sigma=-0.2$, $U=0.5$, $J=U/8$ at half filling and with two
  additional electrons. These are 
ED results on a $\sqrt{8}\times\sqrt{8}$ cluster.\label{fig:sk_pdp0.1_doped} }  
\end{figure}

As it can be observed in Fig.~\ref{fig:sk_U}, the onsite repulsion $U$ enhances the
spin ``striped'' ordering, which is already dominant even at $U$=0 (although in this
noninteracting case a power-law
decay in the spin correlations is expected, rather than genuine long-range order). 
Increasing the Hund's coupling $J$ at a fixed $U$, see Fig.~\ref{fig:sk_J},
produces a similar effect, because it leads to 
larger localized moments, allowing for a stronger overall collective spin 
ordering. However,
for positive $pd\pi$, the diagonal hopping $t_3$ is no
longer strong enough to drive the $(\pi,0)$ order, and the spin
structure factor peaks at $(\pi,\pi)$ instead
(Fig.~\ref{fig:sk_pdpi0.1}). Figure~\ref{fig:sk_J_doped}
shows the spin structure factor for $pd\pi/pd\sigma=-0.2$, $U=0.5$, and
$J=U/8$ when two more electrons are added to half filling. 
The $(0,\pi)$ order is weakened in the doped
system.

In our previous effort,\cite{ours} we investigated the pairing symmetry
for two added electrons
in the region of hoppings $-0.5 \le pd\pi \le -0.2$, varying $U$, and 
for the special case
$J/U=1/4$. The spin of the state with two additional electrons can be
determined by comparing the ground state energy for a total $z$
component of the spin $S_z=0$ to $S_z=1$. If these two energies are
degenerate, the state is a triplet (it was also tested that
the ground state of $S_z=2$ is not degenerate with $S_z=0$ and 1, thus
excluding higher spin states). 
The Hubbard repulsion $U$ was found to drive
the spin of the two electrons added to the half-filled system from
triplet at small $U$ to singlet at larger $U$, for the $pd\pi$'s 
investigated. The critical $U_c$
needed for the transition was found to be the lowest at $pd\pi \approx
-0.2$. This value of $pd\pi$ moreover leads to a Fermi
surface with hole and electron pockets 
similar to that obtained with
band structure after folding [see Ref.~\onlinecite{ours} and Fig.~\ref{fig:fs_pdp-02}].  For large 
$|pd\pi/pd\sigma|$, on the other hand, the Fermi surface has far
larger electron pockets around $(0,\pi)$ and $(\pi,0)$ than those
found in band calculations or experiments, see Fig.~\ref{fig:fs_pdp-05}. 
Consequently, we will mainly focus on $pd\pi/pd\sigma=-0.2$ below.

\begin{figure}
\includegraphics[width=0.45\textwidth]{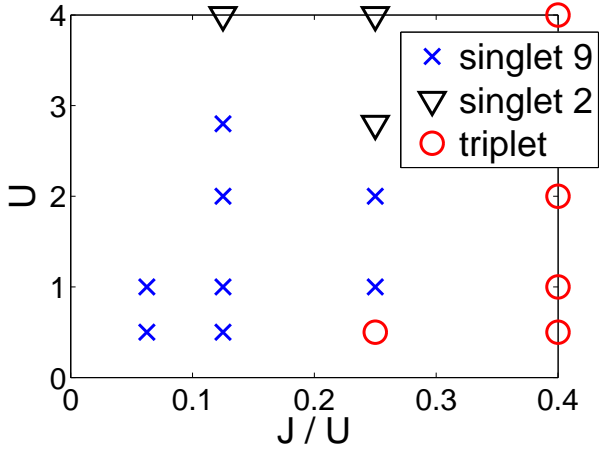}
\caption{(Color online) Dominant pairing tendencies of the ground state corresponding to 
two more electrons
  than half-filling, at $pd\pi$=-0.2. The notation ``singlet 9'' and
``singlet 2'' is explained in the text.
\label{fig:st_J_U_02} 
}  
\end{figure}

In Fig.~\ref{fig:st_J_U_02}, the regions where 
singlet and triplet pairing dominate,   
depending on $U$ and $J$, are shown. (The notation ``singlet 9'' and ``singlet 2''
refer spin-singlet states with $B_{\rm 2g}$ and $A_{\rm 1g}$ symmetry, as
discussed in more detail in the ``Pairing symmetry'' section below as well as in
App.~\ref{app:spm}.)
The trends observed 
for $J/U=1/4$ remain stable for other realistic values of $J$. 
Additionally, qualitatively we have observed
that increasing the Hund coupling $J$ promotes a robust 
triplet pairing. Due to the relation used between $U$, $U'$, and $J$,
two electrons on the same site, but in different orbitals, no
longer feel any Coulomb repulsion for the maximal $J=2U/5$. 
As a consequence, values of 
$J/U \ge 0.4$ are here considered unphysical. 

\subsubsection{Pairing symmetry}

To investigate the symmetry under rotations of the pairing states,
the half-filled ground state is compared to states with two additional electrons. 
The symmetry sector of the half-filled ground state must be contrasted with the symmetry of the
doped state, and the symmetry operation leading from one to the other
gives us an indication for the pairing symmetry. This method was very successful
in establishing the $d$-wave character of the pairing in 
the $t$-$J$ model for the cuprates.\cite{Rev1994} In addition, we also added
a pair of electrons  
with a well-defined symmetry under rotations to the half-filled
ground state, and calculate the overlap between the resulting state and the
ground state obtained for half filling plus two electrons.

From this analysis,
we found that the dominant pairing operator for 
spin-singlet pairs is inter-orbital and given by\cite{ours}
\begin{equation}\label{eq:our_pairing_singlet}
\Delta_9^\dagger = \frac{1}{2N_\textrm{sites}}\sum_{{\bf i}, \alpha, \mu}( 
d^{\dagger}_{{\bf i}, -\alpha, \uparrow}d^{\dagger}_{{\bf i}+\hat\mu,
    \alpha, \downarrow} - 
d^{\dagger}_{{\bf i}, \alpha, \downarrow}d^{\dagger}_{{\bf i}+\hat\mu,
    -\alpha, \uparrow })\;,
\end{equation}
where ${\bf i}=1,\dots,N_\textrm{sites}$ denotes the lattice site, $\hat\mu=\hat x, \hat y$ the
unit vector connecting NN sites, and $\alpha = x, y$ the $xz$ and $yz$ 
orbitals, respectively. This
operator transforms as $B_{\rm 2g}$, and it is $\#9$ in the detailed list provided in
Ref.~\onlinecite{wan}. 
In addition to the $B_{\rm 2g}$ pairing between nearest neighbor sites,
we also find a small overlap for the corresponding $B_{\rm 2g}$
onsite pairing $\#8$ (reaching at most $10\%$ of the
intersite overlap) and some overlap for the NNN $B_{\rm 2g}$
pairing. In contrast to the small onsite contribution, the NNN pairing
is sizable, but its exact strength compared to NN pairing is
difficult to ascertain with the small cluster used.

The only other singlet pairing for which we have found
a substantial overlap is $\#2$ in the above mentioned list, although, as 
discussed below, its region of stability at large $U$ does not have the correct
properties expected for the FeAs new superconductors. This operator is
intra-orbital, has $A_{\rm 1g}$ symmetry, and it is given by
\begin{equation}\label{eq:pairing_singlet_2}
\Delta_2^\dagger = \frac{1}{2N_\textrm{sites}}\sum_{{\bf i}, \alpha, \mu}( 
d^{\dagger}_{{\bf i}, \alpha, \uparrow}d^{\dagger}_{{\bf i}+\hat\mu,
    \alpha, \downarrow} - 
d^{\dagger}_{{\bf i}, \alpha, \downarrow}d^{\dagger}_{{\bf i}+\hat\mu,
    \alpha, \uparrow })\;.
\end{equation}

\begin{figure}
\includegraphics[width=0.45\textwidth]{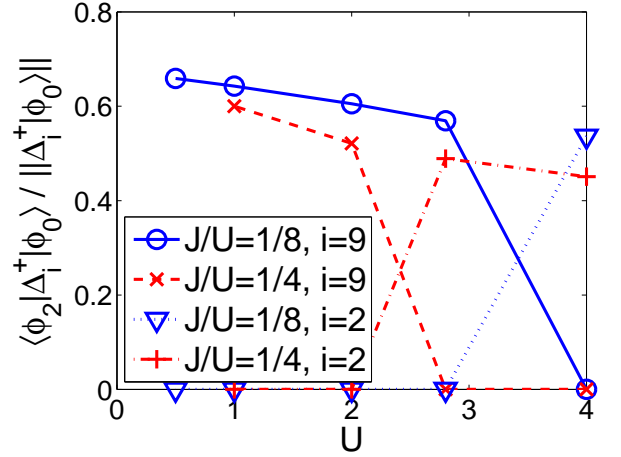}
\caption{(Color online) Overlap between the ground state for half
  filling plus two electrons $|\phi_2\rangle$, and the state obtained by
  applying the pairing operators $\Delta^\dagger_2$ and $\Delta^\dagger_9$
  to the undoped ground  state $|\phi_0\rangle$. At small to
  intermediate $U$, the  inter-orbital singlet pairing operator
  $\Delta^\dagger_9$ with $B_{\rm 2g}$ symmetry ($i=9$) dominates, see
  Eq. (\ref{eq:our_pairing_singlet}). For large $U$, the intra-orbital
  singlet $\Delta^\dagger_2$ ($i$=2) dominates. Results were obtained using $pd\pi/pd\sigma
  = -0.2$, and the ED technique 
on $\sqrt{8}\times \sqrt{8}$ site clusters. \label{fig:overlap-02}}   
\end{figure}

Applying the pairing operators $\Delta_i^\dagger$ to the half-filled
ground state $|\phi_0\rangle$, we find that the resulting vector
$\Delta_i^\dagger|\phi_0\rangle$ has a very small norm
$\lesssim 0.15$ for pairings $i=\#3$, $\#4$, $\#5$ and $\#6$ in the list of 
Ref.~\onlinecite{wan}, while it reaches
$\approx 0.6-0.8$ (depending on $U$ and $J$) for $\#1$, $\#2$, $\#7$,
$\#8$ and $\#9$. Then, at least qualitatively, we 
conclude that only the latter pairs can be created
easily in the half-filled ground state. 
To provide more quantitative information, we then calculate the overlap
between $\Delta_i^\dagger|\phi_0\rangle$ and the ground state found
for half filling plus two additional electrons $|\phi_2\rangle$. We
only find substantial overlaps for the operators $\#9$ ($B_{\rm 2g}$)
and $\#2$ ($A_{\rm 1g}$) given in
Eqs.~(\ref{eq:our_pairing_singlet}) and (\ref{eq:pairing_singlet_2}), 
while $\Delta_7^\dagger|\phi_0\rangle$ is always
orthogonal to the two-electron ground state $| \phi_2\rangle$, at least
for the range of parameters investigated. As it can be observed in
Fig.~\ref{fig:overlap-02}, the $B_{\rm 2g}$ pairing
Eq.~(\ref{eq:our_pairing_singlet}) occurs at small to intermediate
Coulomb repulsion $U$, which is the expected suitable regime to describe non-insulating 
materials with bad metallic properties. 
Only for large $U\gtrsim 2.8$~eV, where a hard gap
in the density of states indicates insulating behavior,~\cite{ours} 
we do find the pairing Eq.~(\ref{eq:pairing_singlet_2}) with $A_{\rm
  1g}$ symmetry. In this regime we find some admixture of the corresponding 
onsite pairing ($\#1$) and the longer-range $A_{\rm 1g}$ NNN pairing which,
as shown in App.~\ref{app:spm}, corresponds to the much discussed $s\pm$ pairing state.
\cite{korshunov,kuroki,mazin,parker} 

Figure~\ref{fig:st_J_U_02} shows more explicitly the regions of dominance of
the states $\#9$ ($B_{\rm 2g}$) and $\#2$ ($A_{\rm 1g}$) in the $U$ vs. $J/U$ plane. 
Thus, the $B_{\rm 2g}$ symmetric operator seems to be the most realistic
in the regime $pd\pi/pd\sigma
\approx -0.2$ and $0.5 \lesssim U \lesssim 1$, which is the appropriate
region of parameters to qualitatively describe the new FeAs-based 
superconductors.\cite{comment}

\subsection{Results with hopping parameters fitted \\
 to band-structure calculations}

\subsubsection{Results at nonzero $J$}

\begin{figure}
\subfigure{\includegraphics[width=0.23\textwidth]{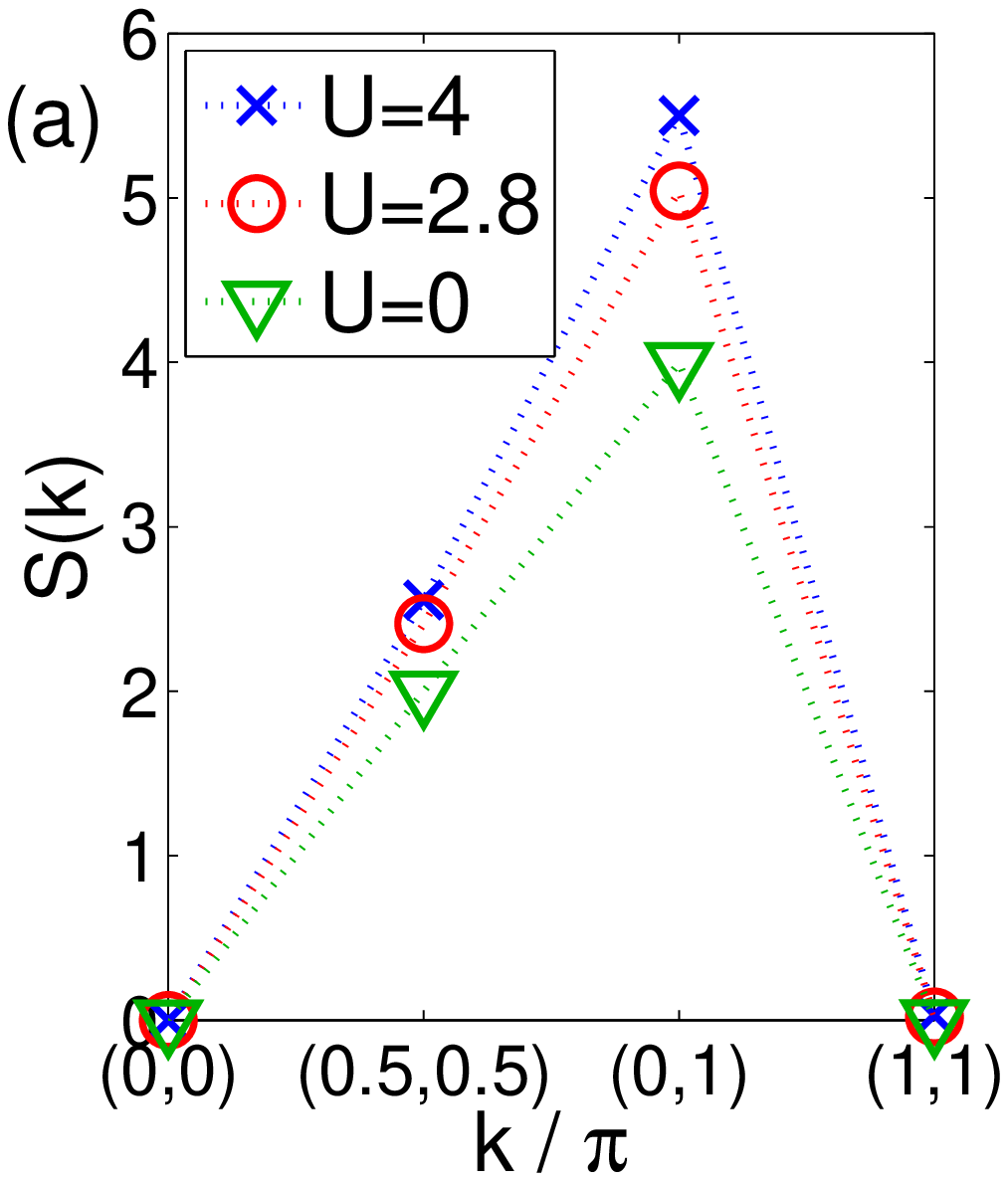}\label{fig:sk_U_sc}}
\subfigure{\includegraphics[width=0.23\textwidth]{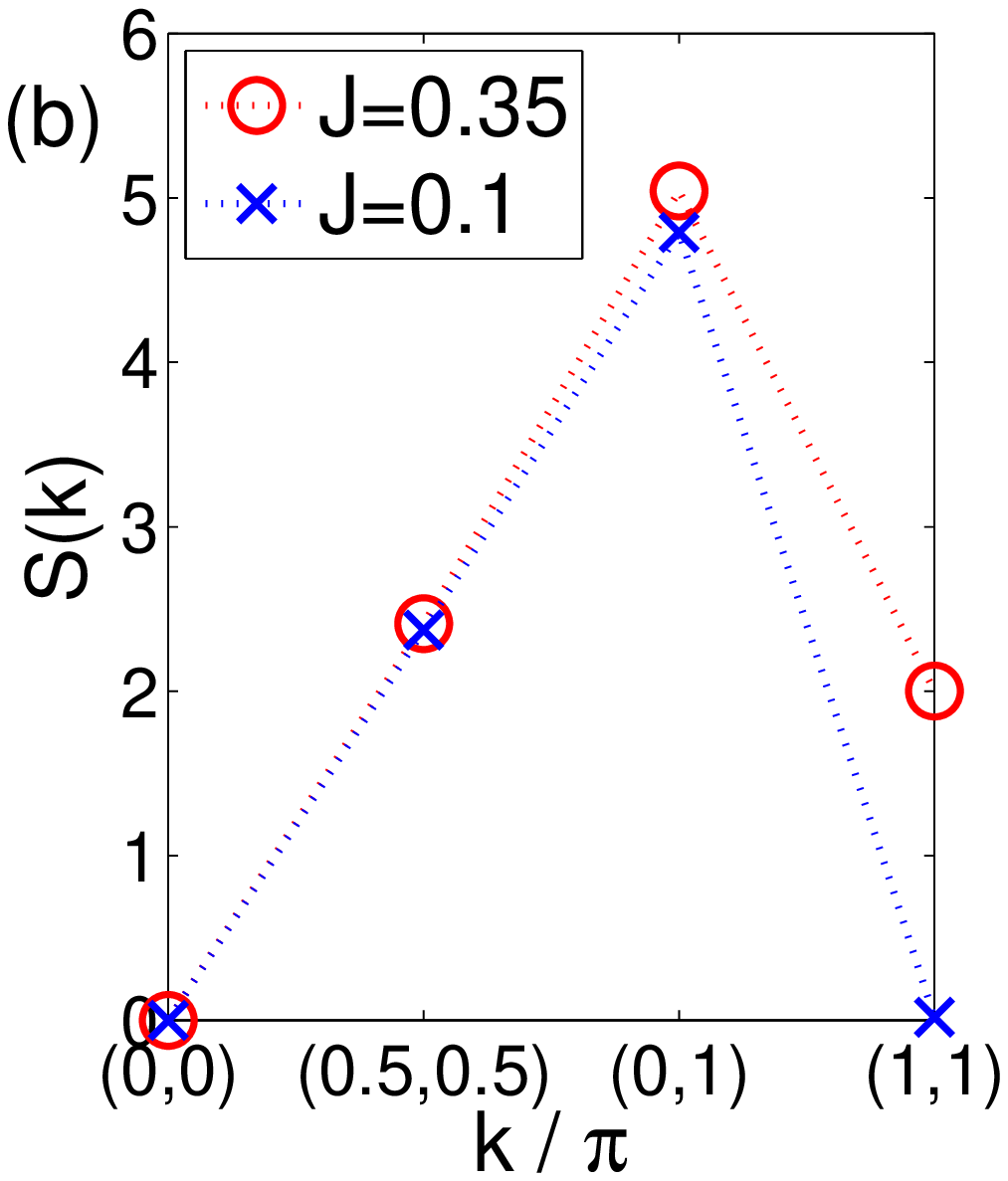}\label{fig:sk_J_sc}}
\caption{(Color online) Spin structure factor $S(k)$  for
  hopping parameters $t_1=-1.0$, $t_2=1.3$, $t_3=t_4=-0.85$ 
(in eV units):\cite{scalapino}  (a) results 
for several values of $U$ and $J/U=1/8$;
(b) results for two values of $J$, with $U=2.8$~eV fixed. 
(a) and (b) were obtained at half filling
  using ED on $\sqrt{8}\times\sqrt{8}$ clusters.\label{fig:sk_U_J_sc} }  
\end{figure}

\begin{figure}
\includegraphics[width=0.23\textwidth]{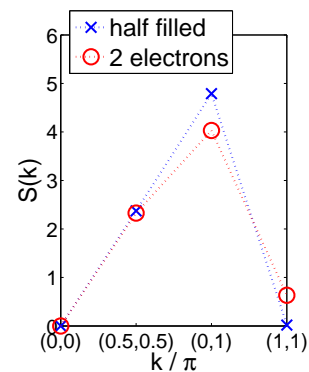}
\caption{(Color online) Spin structure factor $S(k)$  for half filling
  and half filling plus two electrons, using the  hopping parameters $t_1=-1.0$, $t_2=1.3$, 
$t_3=t_4=-0.85$,\cite{scalapino} and $U=2.8$, $J=0.1$ (in eV units).
These are ED results for $\sqrt{8}\times\sqrt{8}$ clusters.\label{fig:sk_sc_2e} }  
\end{figure}

\begin{figure}
\includegraphics[width=0.45\textwidth]{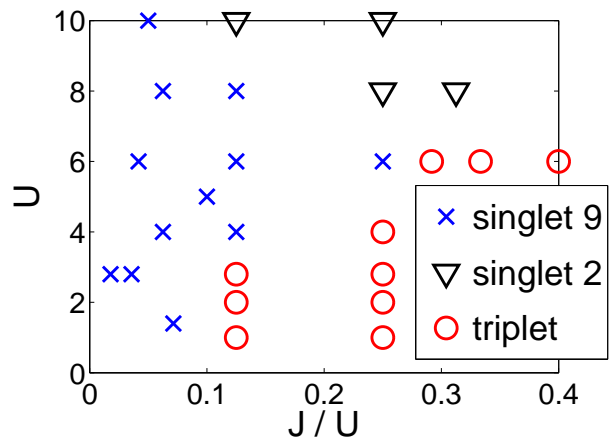}
\caption{(Color online) Dominant pairing tendencies of the ground state for two more electrons
  than half-filling, with hoppings fitted to band-structure
  results.\cite{scalapino} The ``singlet 9'' and ``singlet 2'' notation is
the same as for the case of the Slater-Koster hoppings.
\label{fig:st_J_U_sc} }   
\end{figure}

We have also investigated the two-orbital model using hopping parameters
obtained from a fit to band-structure calculation 
results.~\cite{scalapino} It is interesting to observe that 
this set of parameters also leads to $(0,\pi)$/$(\pi,0)$ antiferromagnetic order at half
filling, see Fig.~\ref{fig:sk_U_J_sc}, which is again enhanced by increasing $U$
at fixed $J$, or increasing $J$ at fixed $U$. As for SK hoppings,
Fig.~\ref{fig:sk_sc_2e} shows that the magnetic order is only slightly reduced by the doping with
two electrons.
In Fig.~\ref{fig:st_J_U_sc}, we report the spin of the state
with two electrons added to the half-filled state and find
qualitatively similar behavior as with the SK approach: $U$
promotes singlet pairing, in the previously discussed 
``\#9'' ($B_{\rm 2g}$) and ``\#2'' ($A_{\rm 1g}$) channels, and $J$
favors triplet pairing.  

We have performed an analogous analysis of the electron pairing as in the previous subsection,
and again find  the inter-orbital singlet operator
Eq.~(\ref{eq:our_pairing_singlet}) to dominate at intermediate $U$ and
$J$. Table~\ref{tab:ov_9_sc} gives the pairing amplitudes for
operators Eqs.~(\ref{eq:our_pairing_singlet}) and (\ref{eq:pairing_singlet_2}),
for several $(U, J)$ parameter sets. As before, the Coulombic parameters were chosen to
give  a spin-singlet state for two electrons added to half filling, but are
expected to be small enough to remain in the metallic regime. As for 
SK hoppings, the pairing symmetry for
these singlet states is $B_{\rm 2g}$, i.e., 
the inter-orbital singlet Eq.~(\ref{eq:our_pairing_singlet}) dominates here as well.

\begin{table}
 \caption{Overlap between the ground state $|\phi_2\rangle$  for two
electrons added to half
  filling  and the states $\Delta^\dagger_i|\phi_0\rangle$ that are
  obtained by applying
  pairing operators Eqs.~(\ref{eq:our_pairing_singlet})
and (\ref{eq:pairing_singlet_2}) to the ground state
  at half filling. Data are for a kinetic energy operator obtained from
  band-structure fitting with  $t_1=-1.0$, $t_2=1.3$, and
$t_3 =t_4=-0.85$ (in eV units).~\cite{scalapino}  \label{tab:ov_9_sc}}
 \begin{ruledtabular}
 \begin{tabular}{cccc}
 U & J & $\langle \phi_2 |\Delta^\dagger_2|\phi_0\rangle$& $\langle \phi_2
|\Delta^\dagger_9|\phi_0\rangle$ \\
\hline
1.4 & 0.10 & 0.00 &0.64\\
2.8 & 0.00 & 0.66 & 0.00\\
2.8 & 0.05 & 0.00 & 0.66 \\
2.8 & 0.10 & 0.00 & 0.66\\
4.0   & 0.50 & 0.00 & 0.65\\
10.0 & 1.25 & 0.58 & 0.00 \\
 \end{tabular}
 \end{ruledtabular}
 \end{table}

Inter-orbital pairing is favored over intra-orbital pairing by
the kinetic energy, because the inter-orbital Coulomb repulsion $U'$ is
weaker than the repulsion $U$ within the orbitals. To test this
assumption, we analyze pairing amplitudes for the special case $J=0$ but
instead of using $U=U'$, we use $U'<U$,
i.e., we deviate from the relation $U'=U-2J$. The resulting
pairing symmetry is still the same $B_{\rm 2g}$ inter-orbital singlet
Eq.~(\ref{eq:our_pairing_singlet}), as deduced from the amplitudes
on Tab.~\ref{tab:ov_9_sc_Up}.  

\begin{table}
 \caption{Overlap $\langle\phi_2|\Delta^\dagger_i|\phi_0\rangle$ as in
   Tab.~\ref{tab:ov_9_sc}, but with $J=0$. The coupling $U'$ is different
   from $U$, and the hoppings were $t_1=-1.0$, $t_2=1.3$, and 
$t_3 =t_4=-0.85$ (in eV units).
\label{tab:ov_9_sc_Up}}
 \begin{ruledtabular}
 \begin{tabular}{cccc}
 U & U' & $\langle \phi_2 |\Delta^\dagger_2|\phi_0\rangle$ & $\langle \phi_2 |\Delta^\dagger_9|\phi_0\rangle$ \\
\hline
2.8 & 2.8 & 0.66 & 0 \\
2.8 & 2.6 & 0 & 0.66 \\
2.8 & 1.4 & 0 & 0.62 \\
 \end{tabular}
 \end{ruledtabular}
 \end{table}

\subsubsection{Results at $J$=0 and $U=U'$}

Finally, let us discuss the special case $J=0$, $U=U'$ for $U=2.8$ (in eV units), 
which are the couplings
used in Ref.~\onlinecite{scalapino}. In contrast to
$J\gtrsim 0.05$ and $U' \lesssim 2.6$, we here find that the $A_{\rm 1g}$
intra-orbital singlet Eq.~(\ref{eq:pairing_singlet_2}) has the lowest energy, i.e. the same
pairing as observed at large $U$ for both the fitted and the SK hoppings.  
However, at $U=U'=2.8$ and $J=0$, the second lowest  eigenstate 
is almost degenerate with the ground state, and it 
gives an overlap with Eq.~(\ref{eq:our_pairing_singlet}), i.e. pairing $\#9$. 
The third state at these
couplings corresponds to
the intra-orbital pairing $\#7$ ($B_{\rm 1g}$). Table \ref{tab:ov_sc_J0} 
contains the explicit numbers
 for energies and overlaps. The near degeneracy of these states does not allow
us to reach a clear conclusion for the $J=0$ and $U=U'$ case, which appears to 
be singular, since small modifications away from our results, such as 
increasing lattice sizes, may change the relative order
of the competing states.

\begin{table}
\caption{Energy of the lowest eigenstates for two electrons and $J=0$, $U=U'=2.8$,
and pairing operators giving the largest overlap when applied to the
half-filled state. Hoppings are those from band-structure fitting.\label{tab:ov_sc_J0}}
\begin{ruledtabular}
\begin{tabular}{cccc}
& energy & pairing & $\langle \phi_2 |\Delta^\dagger_i|\phi_0\rangle$ \\
\hline
1 & -8.45322 & $\#2$ & 0.66 \\
2 & -8.45150 & $\#9$ & 0.66 \\
3 & -8.4132  & $\#7$ & 0.67 \\
\end{tabular}
\end{ruledtabular}
\end{table}

\subsubsection{Competing magnetic states at half-filling}

\begin{table}
 \caption{Four-spin correlations $\langle v_x|v_x\rangle$ and $\langle
 v_y|v_x\rangle$ for two parameter sets using the full two-orbital model at
the values of $U$ and $J$ indicated, on an 8-site cluster. For comparison,
results using  a perfect
 Ising-like $(0,\pi)+(\pi,0)$ state on 8 sites are also shown.\label{tab:4spin}}
 \begin{ruledtabular}
 \begin{tabular}{ccccc}
  & Ising & $U=4$ & $U=1$& $U=0.5$\\
  &       & $J=1$ & $J=0.25$& $J=0.0625$\\
\hline
$\langle v_y|v_x\rangle$ & -1  & -0.64 & -0.1684 & -0.07776 \\
$\langle v_x|v_x\rangle$ & 0.0625  & 0.02735 &  0.02856 & 0.0225\\
 \end{tabular}
 \end{ruledtabular}
 \end{table}

The magnetic phase diagram for the
present model was also studied using a mean-field approximation.\cite{lorenzana} At half
filling, it was claimed that the Coulomb repulsion rather than stabilizing 
a state with
$(\pi,0)/(0,\pi)$ spin-stripe order, induces an ``orthomagnetic'' (OM)
ordering where NN spins are at
right angles. The spin structure factor for this phase is still
peaked at $(0,\pi)$ and $(\pi,0)$. We have tried to address 
this issue with the ED technique on the 8-site cluster. Unfortunately, several 
observables are expected to give similar results for the two states. Apart
from having similar spin structure factors, diagonal NNN spin-spin
correlations are also negative in both states. Moreover, even the
expectation value for NN correlations vanishes in both cases: in the
OM state because the NN spins are at $90^\circ$, while in the stripe state the cancellation
occurs because the small cluster
ground state $|\phi_0\rangle$ contains both $(0,\pi)$ and $(\pi,0)$ configurations 
with equal weight and
NN correlations $\langle \phi_0 | v_x\rangle$ and $\langle \phi_0 |
v_y\rangle$ average out, where
\begin{align}
|v_x\rangle &= \frac{1}{N_\textrm{sites}}\sum_{\bf i} {{\bf S}_{\bf i}}\cdot{{\bf S}_{{\bf i}+{\hat x}}}|\phi_0\rangle\;,\\|v_y\rangle &= \frac{1}{N_\textrm{sites}}\sum_{\bf i} {{\bf S}_{\bf i}}\cdot{{\bf S}_{{\bf i}+{\hat y}}}|\phi_0\rangle\;.
\end{align}
Our numerical ED results indeed give very small negative values for the NN spin correlation.
However, we expect the two states to lead to different results for $\langle
v_x|v_x\rangle$ and $\langle v_y|v_x\rangle$: $\langle
v_x|v_x\rangle$ is expected to be positive and $\langle
v_y|v_x\rangle$ negative in the spin-striped phase, while both should be
zero or very small in the OM phase. Table~\ref{tab:4spin} shows the results for
strong ($U=4$), intermediate ($U=1$), and weak ($U=0.5$) on-site Hubbard repulsion.
For $U=4$, we clearly find $\langle
v_y|v_x\rangle < 0$ and $\langle v_x|v_x\rangle>0$. While these
numbers become weaker for the less spin ordered states at smaller $U$, they
are still consistent with the $(0,\pi)/(\pi,0)$ ordering. For
comparison, we also include results obtained for a linear combination
of states with perfect Ising-like $(0,\pi)$ and $(\pi,0)$
order in the $z$-direction. These results are useful to judge the
expected order-of-magnitude values for the correlations investigated. The
numbers on Table VI show that the numerical results for the two-orbital model
are compatible with those of the Ising spin-stripe state, particularly
considering that quantum fluctuations will reduce the spin correlations of
such a state. As a consequence, our present investigations favor
the spin-stripe magnetic state, although further work is needed to fully
confirm these conclusions.

\section{Discussion of nodal structure in the mean-field approximation}

In this section, the results of 
a pairing mean-field analysis of the two-orbital 
Hamiltonian will be discussed. The numerical results of the previous sections and 
experimental data will be used to guide this mean-field approximation.
Experiments indicate that the pairs in the 
Fe-based superconductors are spin singlets \cite{Grafe,matano,kawabata}   and 
our previous numerical results did
provide a dominant spin-singlet pairing operator, as discussed in the previous section. 
It is also important to notice 
that the pairing operator that we obtained mixes different 
orbitals. A numerical study of the orbital composition of the bands that 
determine the FS in our two-orbital model indicates that 
the bands that constitute the pockets are an 
admixture of $xz$ and $yz$ orbitals. Thus, it is not surprising that the 
dominant pairing operator creates pairs made of electrons in different orbitals.
One of the main goals of the analysis discussed below will be
to find out the nodal structure of the mean-field Hamiltonian.

\subsection{Location of the Nodes}

\subsubsection{Reminder of one-band model results.}
For the simple case of $d$-wave superconductivity in a single-orbital model,
characterized by the dispersion relation $\xi({\bf k})=-2t(\cos k_x+\cos k_y)-4t'\cos k_x
\cos k_y-\mu$, the gap function is given by 
$\Delta_k=\Delta(\cos k_x-\cos k_y)$. In this case, the mean-field Hamiltonian reduces
to a 2$\times$2 
matrix linking $\bf k$ with -$\bf k$ which is simply given by
\begin{equation}
H_{\rm MF}=
 \left(\begin{array}{cc  }
\xi({\bf k}) & \Delta_k \\
\Delta_k  &-\xi({\bf k})  
\end{array} \right).
\label{40}
\end{equation}
To obtain the position of the nodes in the gap, we merely need to find the 
values of $k_x$ and $k_y$ where the eigenvalues of the matrix 
Eq.~(\ref{40}) are zero. These are the same values that solve the equation 
$det (H_{\rm MF})=0$, i.e.,
$\xi({\bf k})^2+\Delta_k^2=0,$
which is satisfied only if each term vanishes independently. This occurs at the
points where the non-interacting Fermi surface described by $\xi({\bf k})=0$ 
intersects the diagonal 
lines along which $\Delta_k=0$, i.e., $k_x=k_y$ and $k_x=-k_y$.
This procedure establishes the well-known location of the four $d$-wave nodes of a single-band
model.

\subsubsection{Nodes in a two-orbital model}

For a system with two orbitals, we will proceed in an analogous manner as for
one orbital. The
MF Hamiltonian matrix in the basis 
$(d^{\dagger}_{{\bf k},x,\uparrow},d^{\dagger}_{{\bf k},y,\uparrow},
d_{{\bf -k},x,\downarrow},d_{{\bf -k},y,\downarrow})$ is now given by the 4$\times$4 matrix
\begin{equation}
H_{\rm MF}=
 \left(\begin{array}{cccc}
\xi_{xx}  & \xi_{xy}               &     0         & \Delta_k \\
\xi_{xy}  & \xi_{yy}               &     \Delta_k  & 0 \\
0         & \Delta_k               & -\xi_{xx}     & -\xi_{xy}  \\
\Delta_k  & 0                      & -\xi_{xy}     & -\xi_{yy}  
\end{array} \right) .
\label{42}
\end{equation}
The matrix elements can be obtained by Fourier transforming the tight-binding
Hamiltonian $H_{\rm TB}$ given in 
Eq.~(\ref{20}). We obtain:
\begin{eqnarray}
\xi_{xx}&=&-2t_2\cos k_x-2t_1\cos k_y-4t_3\cos k_x\cos k_y-\mu, \nonumber \\
\xi_{yy}&=&-2t_1\cos k_x-2t_2\cos k_y-4t_3\cos k_x\cos k_y-\mu, \nonumber \\
\xi_{xy}&=&-4t_4 \sin k_x \sin k_y,
\label{45}
\end{eqnarray}
\noindent and
\begin{equation}
\Delta_k=V(\cos k_x+\cos k_y),
\label{46}
\end{equation}
\noindent where $V$ is the strength of the pairing interaction.

Notice that we can also work in the basis in which $H_{\rm TB}$ is diagonal. In
this basis, which is expanded by $(c^{\dagger}_{k,1,\uparrow},c^{\dagger}_{k,2,\uparrow},
c_{-k,2,\downarrow},c_{-k,1,\downarrow})$, $H_{\rm MF}$ becomes $H'_{\rm MF}=U^{-1}
H_{\rm MF}U$ given by:
\begin{equation}
H'_{\rm MF}=
\left(\begin{array}{cccc}
\epsilon_1 & 0               &     V_{\rm B}  & V_{\rm A} \\
0 & \epsilon_2               &     -V_{\rm A}  & V_{\rm B} \\
V_{\rm B}  & -V_{\rm A} & -\epsilon_2 & 0  \\
V_{\rm A}  & V_{\rm B} & 0 & -\epsilon_1  
\end{array} \right),
\label{46a}
\end{equation}
\noindent where $V_{\rm A}$ and $V_{\rm B}$ are given by:
\begin{equation}
V_{\rm A}=2uv\Delta_k, \\
\label{47}
\end{equation}
\begin{equation} 
V_{\rm B}=(v^2-u^2)\Delta_k,
\label{48}
\end{equation}
\noindent and $u$ and $v$ are the elements of the change of basis matrix $U$ 
given by
\begin{equation}
U=
 \left(\begin{array}{cccc}
u & v               &     0  & 0 \\
v & -u               &     0  & 0 \\
0  & 0 & v & u  \\
0  & 0 & -u & v  
\end{array} \right) ,
\label{49}
\end{equation}
\noindent with $U^{-1}=U^T$.
Remember that $V_{\rm A}$, $V_{\rm B}$, $u$, and $v$ are all functions of the 
momentum $k$, and $u^2+v^2=1$. 
\begin{figure}[thbp]
\begin{center}
\includegraphics[width=8cm,clip,angle=0]{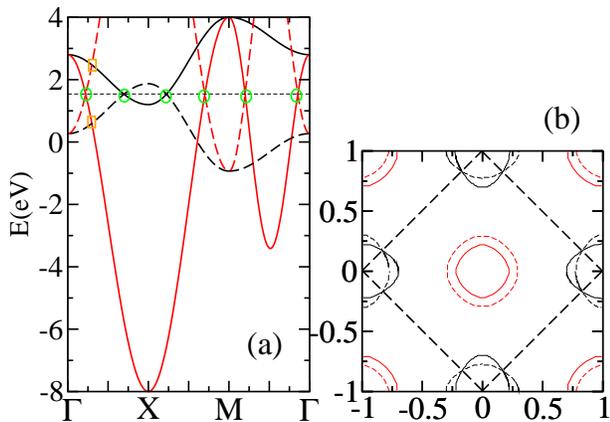}
\vskip 0.3cm
\caption{(Color online)
(a) Energy vs. momentum for the non-interacting 
tight-binding Hamiltonian
Eq.~(\ref{20}) using the hoppings from 
band-structure calculations,\cite{scalapino} 
$t_1 = -1$,$t_2 = 1.3$,
and $t_3  =t_4 = -0.85$ (in eV units). These non-interacting results are shown
as continuous lines. Also shown are the additional Bogoliubov bands produced by 
the pairing interaction considered in this work (dashed lines). The 
circles (boxes) indicate the bands that contribute electrons to the intraband 
(interband) pairs at different 
locations close to the FS (for a discussion see text). 
(b) Fermi surface for the corresponding non-interacting half-filled system.}
\label{dougshadow}
\end{center}
\end{figure}

\begin{figure}[thbp]
\begin{center}
\includegraphics[width=8cm,clip,angle=0]{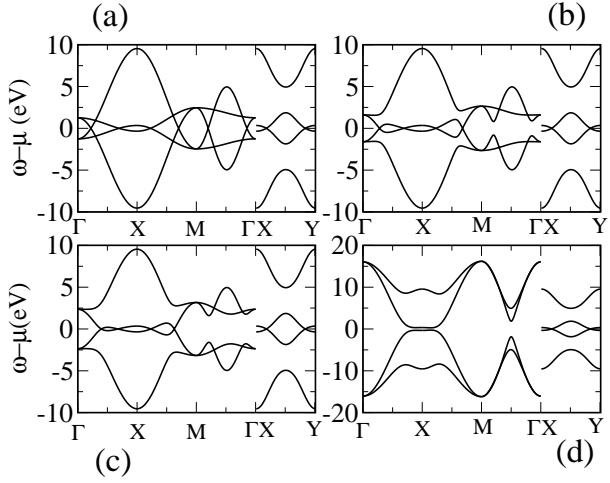}
\vskip 0.3cm
\caption{(Color online)
(a) Energy vs. momentum for the mean-field
Hamiltonian
using 
$t_1=-1$,$t_2=1.3$,
and $t_3=t_4=-0.85$ (in eV units).\cite{scalapino} 
The Bogoliubov bands produced by 
the pairing interaction considered in this work are also shown, with
equal intensity.
The four panels correspond to four
different values of the pairing attraction: (a) $V=0$,
(b) $V=0.5$, (c) $V=1$, and (d) $V=8$}
\label{mfdoug}
\end{center}
\end{figure}

Now let us discuss the physical meaning of $V_{\rm A}$ and $V_{\rm B}$.
According to Eq.~(\ref{46a}), $V_{\rm A}$ is the intraband pairing for
band $1$, i.e., the band with the highest energy (electron band), while the 
intraband pairing
for band $2$ (hole band) is $-V_{\rm A}$. Thus, there is a relative 
phase $\pi$ 
between the two intraband order parameters. 
In the standard BCS studies for multiband models, it is expected 
that pairs are formed by electrons in the same band.\cite{Suhl} In the two 
orbital model, as just discussed, we found intraband pairing 
but we also obtain {\it interband pairing} with strength $V_{\rm B}$.
The possibility of interband pairing has been considered previously
in several contexts: (i) Possibility of $T_c$\cite{kumar}, (ii) high $T_c$ 
cuprates,\cite{kheli} and (iii) heavy fermion systems,\cite{khomskii} where it 
was shown that
if  two bands are very close to each other in the 
vicinity of the Fermi level, interband pairing can 
occur. 
The weaker the pairing potential the closer to the Fermi surface the 
two bands have to be.
When long range pairing develops the Brillouin zone gets folded and, as a 
result, the total number of bands doubles. In this representation, which 
arises by diagonalizing Eq.(\ref{42}) with $\Delta_k=0$, the two-orbital model 
has the dispersion shown in Fig.~\ref{dougshadow} where each band (panel (a))
and FS (panel (b)) is 
represented with a different color and the folded (unfolded) portions with 
dashed (continuous) lines. It can be seen from 
panel (a) that at the Fermi level there is 
only intraband crossing indicated by green circles. However, there is also
interband crossing, indicated by the orange boxes, above the Fermi 
energy. The previous numerical results appear to indicate that the effective coupling 
is sufficiently strong as to produce interband pairing. In Fig.~\ref{mfdoug} 
it can be seen that even a small $V$ opens a gap between the two bands that 
cross away from the FS. In multiorbital models the opening of these gaps can
lower the overall energy.\cite{Yu} Within the 
standard BCS approach this 
result may appear counterintuitive 
and it could be an artifact of the two-orbital model
or of the small system size that we have considered. However, 
there are clear indications that most of the FS in the 
five-orbital model have character
$d_{\rm xz}$ and $d_{\rm yz}$\cite{fang2,vero} and, 
in such a case, the only possible 
pairing operators
that respect the symmetry of the FeAs planes are those that have been 
considered in our calculations and others.\cite{wan} 
In fact, none of the 16 pairing
operators that combine these two orbitals leads to a purely intraband pairing
interaction. This, of course, could be an indication that other orbitals have to
participate in the model but we believe that it is still 
instructive to consider the nodal structure that results from the pairing 
operator that was favored numerically within the two-orbital model and
attempt to compare the results with experiments.

The existence and position of nodes in the resulting 
mean-field band structure can be found by requesting that det$(H'_{\rm MF})=0$, 
as in the one-orbital case. 
From Eq.~(\ref{46a}), we obtain the following equation:
\begin{equation}
V_{\rm A}^2(V_{\rm A}^2+2V_{\rm B}^2+\epsilon_1^2+\epsilon_2^2)+(\epsilon_1\epsilon_2+V_{\rm B}^2)^2=0.
\label{50}
\end{equation}
This equation is satisfied in two possible ways:
\vspace{0.1cm} 

(1) First, a solution can be found if 
$V_{\rm A}=V_{\rm B}=0$, and 
$\epsilon_1=0$ or $\epsilon_2=0$. Thus, this
condition for nodes is satisfied if the lines 
where $\cos k_x+\cos k_y=0$, namely the lines where 
$V_{\rm A}$ and $V_{\rm B}$ vanish, intersect any of the 
non-interacting Fermi surfaces determined by the points where $\epsilon_1=0$
or $\epsilon_2=0$. It is clear that the line $\cos k_x + \cos k_y=0$ intersects
each of the four electron-pocket Fermi surfaces in two points per pocket 
(see Fig.~\ref{dougshadow}b and Fig.~1e).
This means that nodes will appear only in the electron pockets, not in the hole
pockets. 
These are the nodes that arise from a simple extrapolation of the
reasoning used to find nodes in the one-orbital model, namely by finding
the intersections of the non-interacting Fermi surface with the trigonometric
function, in this case $\cos k_x + \cos k_y$, 
contained in the $\Delta_k$ gap function. The position of the
nodes in the electron pockets upon folding of the Brillouin zone (see Fig.~1d
and Fig.~\ref{dougshadow}b)
is at the points in $k$-space where the two electron pockets intersect 
each other.

Notice that the existence of these nodes does not depend on 
the value of $V$.
They will always be present as it can be seen in Fig.~\ref{mfdoug} where the 
nodes along the $X-Y$ direction appear in all the panels varying $V$. 

\vspace{0.1cm}

(2) However, the two-orbital nature of this problem leads to the possibility of
additional nodes in unexpected locations. This can be understood 
by realizing that Eq.~(\ref{50})
can also be satisfied if $V_{\rm A}=0$ and
\begin{equation}
V_{\rm B}^2=-\epsilon_1\epsilon_2.
\label{51}
\end{equation}
According to the expression for $V_{\rm A}$ in 
Eq.~(\ref{47}), and assuming that $\Delta_k$ is non-zero (if it is zero we
recover the nodes already described in (1) above), then the condition $V_{\rm A}=0$ is
satisfied 
if $u=0$ or $v=0$. Due to the normalization $u^2+v^2=1$,
when $u=0$ then it must occur that $v=1$, and {\it vice versa}. Introducing these
values of $u$ and $v$ in Eq.~(\ref{49}), it can be shown that 
the condition that $H'_{\rm MF}=U^{-1}
H_{\rm MF}U$ is diagonal is satisfied only if $\xi_{xy}=0$. 
According to Eqs.~(\ref{45}), for
$\xi_{xy}$ to vanish it is necessary to have $k_x=0$ or
$\pi$, or $k_y=0$ or $\pi$. Then, new nodes could be expected 
along these horizontal or vertical  lines in the Brillouin zone.
Since the product of the two energies $\epsilon_1\epsilon_2$ has to be
negative (i.e. the energies cannot vanish, otherwise we recover (1)), 
the nodes, if they exist, 
will appear {\it in between} the hole
and electron pockets at locations in $k$-space that do 
$not$ belong to the original tight-binding Fermi
surface. To understand this interesting result, consider the example of $k_x=0$.
For this special case we obtain,
\begin{eqnarray}
V_{\rm B}^2&=&V^2(1+\cos k_y)^2, \nonumber \\
\epsilon_1&=&-2t_2-2t_1\cos k_y-4t_3\cos k_y-\mu, \nonumber \\
\epsilon_2&=&-2t_1-2t_2\cos k_y-4t_3\cos k_y-\mu.
\label{54c}
\end{eqnarray}
Replacing Eqs.~(\ref{54c}) 
in Eq.~(\ref{51}), a quadratic equation is obtained that allows
us to find the values of $\cos k_y$ where nodes should appear. Depending on the
specific 
values of $V$, the hopping amplitudes, and $\mu$, the quadratic equation 
can have two
solutions (meaning that two nodes appear along the $k_x=0$ axis between the
hole and electron pockets), 
or one solution (meaning just one node), or no solution at all
(indicating no extra nodes). Thus, once the folding and rotation of 
the FBZ is performed, nodes can appear along the diagonals of the BZ in Fig.~1c for
particular values of the parameter in the model.\cite{another}

\begin{figure}[thbp]
\begin{center}
\includegraphics[width=8.5cm,clip,angle=0]{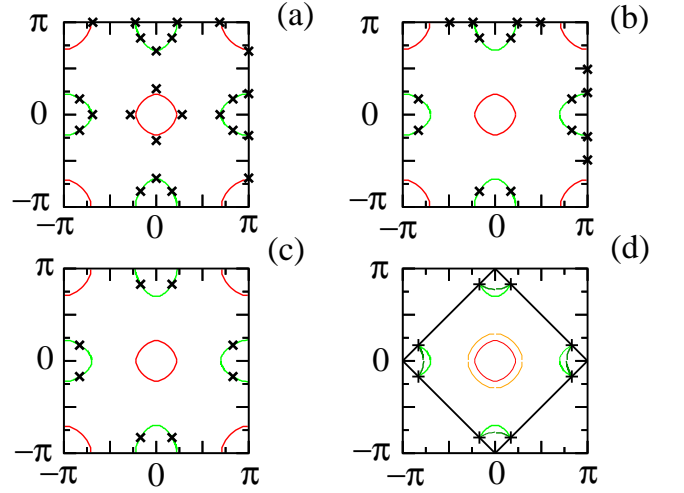}
\vskip 0.3cm
\caption{(Color online) Examples showing the positions of the nodes for
the two-orbital model, using the standard pairing mean-field approximation.
With solid lines are shown the hole pockets (red) and electrons pockets (green), 
namely the Fermi surface in the non-interacting limit $V = 0$.
The black crosses are the nodes. As example, we use the case of the hopping
amplitudes derived from fits with band-structure calculations.\cite{scalapino}
(a) corresponds to the ``weak'' pairing regime, $V=0.5$, showing the existence of many
nodes, both at the electron pockets as well as near the hole pockets. (b) is
the example of $V = 2.0$ where the number of nodes has been reduced
compared with (a). (c) is the ``intermediate'' $V=3.5$ regime ($V$ is still substantially
smaller than the bandwidth) that provides the smaller
number of nodes (a total of 8), 
all being located at the non-interacting electron pockets.
(d) is the folded Fermi surface corresponding to case (c) for better comparison
with experiments and band calculations.
}
\label{nodes}
\end{center}
\end{figure}

A variety of examples obtained numerically illustrate
this nontrivial nodal structure, as shown in Fig.~\ref{nodes}: 
(a) at weak $V$, several
nodes are found either at or close to both the hole and electron pockets. In view of
recent photoemission experiments reporting the absence of nodes at the hole pockets (see
discussion below), 
this regime is unlikely to be realized experimentally. 
(b) is obtained increasing $V$: in this case the number of nodes has decreased.
In addition to those coming from solution (1) in the previous discussion, all at the
electron pockets, still solution (2) provides some nodes at the boundaries of the Brillouin
zone in this regime. (c) is the most canonical result, obtained at intermediate $V$, with the nodes only
appearing in the electron pockets where $\cos k_x+\cos k_y=0$ intersects the original Fermi surface.
Both in (b) and (c) there are no nodes in the $\Gamma$ centered hole pocket even for this
$B_{\rm 2g}$ state.
(d) provides the results of (c) but in the folded zone for comparison with experiments.

Band dispersions obtained with mean field results are also shown in 
Fig.~\ref{mfdoug}. We observe how the nodes along the $\Gamma-X$ and 
$X-M$ directions get closer to each other as $V$ increases from 0 to 0.5 and 
to 1; and how they have disappeared for $V$=8. 
It is also interesting to see how the crossing of different
bands, indicated by the orange squared 
boxes in Fig.~\ref{dougshadow}~(a), is replaced by a gap as soon as $V$ is finite
(see panel (b) in Fig.~\ref{mfdoug}) which appears to be the effect of the interband interaction.

\begin{figure}[thbp]
\begin{center}
\includegraphics[width=8.5cm,clip,angle=0]{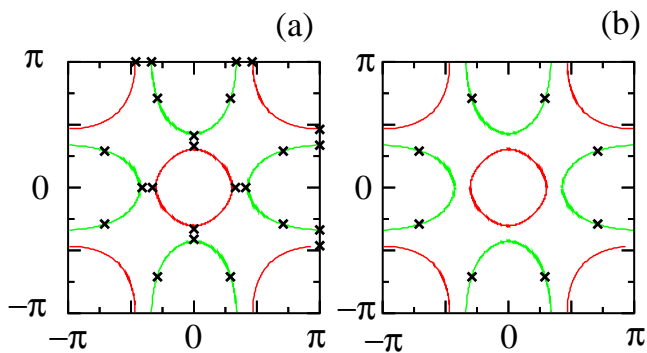}
\vskip 0.3cm
\caption{(Color online) 
Nodal structure for SK hoppings
with $(pd\pi)=-0.2$. (a) is in the weak coupling regime, $V = 0.05$, and shows
many nodes as in Figs.~\ref{nodes}~(a,b). (b) is the intermediate coupling regime, $V= 0.5$, where the
number of nodes is the minimal (8), similarly as in Figs.~\ref{nodes}~(c,d).
}
\label{nodes2}
\end{center}
\end{figure}

For completeness, in Fig.~\ref{nodes2} we provide the nodal structure for the case of
hoppings obtained from the Slater Koster approximation, that gives large  pockets in the
band-structure calculations. In the weak coupling case, (a), once again several nodes are obtained.
This regime appears unrealistic. Increasing $V$, panel (b) shows that the nodes only 
remain in the electron pockets, as found before in Figs.~\ref{nodes}~(c,d).

In addition to the analytic discussion, we have
also searched numerically, using a large lattice 200$\times$200 in $k$-space, 
for the zero eigenvalues of the original
matrix Eq.~(\ref{42}). These numerical results
are in excellent agreement with the analytic discussion, thus
showing 
that the nodal structure of the two-orbital mean-field pairing
Hamiltonian has been properly obtained.

\subsubsection{Comparison with ARPES experiments}

How do these theoretical calculations based on the two-orbital model 
compare with angle-resolved photoemission experiments for the Fe pnictides? 
In Ref.~\onlinecite{arpes}, ARPES results were
presented with the focus of the effort on the hole pockets at $\Gamma$. It was concluded that
nodes were not observed in those hole pockets. This result is compatible with our
$B_{\rm 2g}$ state since nodes do not appear on the hole pockets, but instead 
on the electron pockets, at least at intermediate values of the attraction $V$.
In Ref.~\onlinecite{arpes2},
a similar conclusion was reached but in this case the electron pockets were also studied.
However, in that effort the 122 materials for which the FS depends on $k_z$ were analyzed, 
and only a few cuts in momentum-space were 
investigated. 
Recent ARPES experiments
that suggest a short Cooper-pair size\cite{arpes22} would suggest that the regime of
large $V$ in our study of the nodal structure is the most realistic, thus clearly locating
all the nodes in the electron pockets.
Then, a more detailed ARPES analysis
would be needed to fully conclude that there are no nodes 
in this system, particularly in the electron pockets.\cite{arpes21} 
Other recent ARPES experiments 
have shown a variety of interesting aspects, such as substantial differences with band
structure calculations,\cite{arpes3} that also need to be incorporated in future theoretical
studies.

The information provided in this manuscript for the actual 
location of the nodes for the $B_{\rm 2g}$ state will help to guide
future ARPES experiments. In view of the
several other experimental investigations that have reported 
nodes in the Fe-based 
superconductors,\cite{nodal1,nodal2,nodal3,Ahilan,nakai,Grafe,Y.Wang,matano,mukuda,millo,wang-nodes}
we believe this issue is still open and needs further research to arrive to a final conclusion. If
future experimental work 
clearly proves that there are no nodes in the new Fe-based superconductors, not only in the hole 
pockets but more importantly in the electron pockets,
then it will be concluded
that the two-orbital model used here will not be 
sufficient to properly describe this family of materials,
and more orbitals will be needed.\cite{foot}

\section{Conclusions}

In this manuscript, we have studied some properties of a two-orbital approach for
the new Fe-based superconductors. It is important to find out the minimal model
capable of reproducing the basic physics of these materials. By studying
a relatively simple model, considerable insight could be reached on the inner mechanisms
that cause magnetism and superconductivity in these compounds. While models with more than
two orbitals would certainly be more accurate, the difficulty in extracting reliable
numbers from the models grows fast with the number of orbitals.

Here we have shown that the magnetic properties of the undoped parent compound
are properly reproduced by a simple two-orbital model: spin stripes are 
obtained in agreement with neutron scattering experiments. 
 Regarding electron doping, here we follow the same approach as 
for the cuprates: it is expected
that the pairing channel will be unveiled by simply studying the symmetry properties of
the state of two electrons added to the half-filled ground state. This approach worked
for the models for Cu-oxides superconductors, leading to the $d$-wave state prediction. 
Within this assumption, the spin-singlet pairing state that dominates in the 
phase diagram at
realistic values of the Hubbard repulsion $U$ is found to transform
according to the $B_{\rm 2g}$ representation of the lattice symmetry group. 
At large Coulomb repulsion $U$, too large to describe the metallic state of 
the undoped compound, we found that the relative symmetry of the undoped and
electron doped ground states is the same and, thus, they are connected by a
pairing operator that transforms according to $A_{\rm 1g}$. 
We showed that the NNN pairing operator with this symmetry is the
``$s\pm$'' state and that this state, according to our numerical calculations, 
prevails only in an unphysical regime of parameters. On the other hand, for
a robust electron-electron effective attraction to form Cooper pairs, 
assumption 
compatible with the conclusions of recent ARPES experiments, 
the $B_{\rm 2g}$ pairing state found
for realistic $U$ has nodes only in the electron pockets.
All our main conclusions do not depend
qualitatively on the set of hopping amplitudes used: our results appear to be
representative of the two-orbital framework in general and not merely of a particular
model with particular couplings.
Thus, a conclusion of our study is that more refined ARPES experiments in the
superconducting state are needed to
analyze the possible existence of nodes in the electron pockets. 
These future 
experiments will provide crucial information to guide the theoretical search for the
minimal model that captures the physics of the Fe pnictides.


\section{Acknowledgments}
The authors thank E. Arrigoni and D. Scalapino for useful discussions.
This work was supported by the NSF grant DMR-0706020 and the
Division of Materials Science and Engineering, U.S. DOE, under contract
with UT-Battelle, LLC.

\appendix 
\section{Effective Interaction that Generates the $B_{\rm 2g}$ Pairing Operator}
\label{app:1_eff}
 
The mean-field superconducting state discussed before
could originate from an effective
attractive density-density interaction dynamically generated in
the original Hamiltonian, or induced
by particular phononic modes if an electron-phonon coupling is incorporated. The form
of this attraction is:
\begin{equation}
H_{\rm attr}=-V\sum_{{\bf i},\mu,\alpha,\sigma}n_{{\bf i},\alpha,\sigma}
n_{{\bf i}+\mu,-\alpha,-\sigma},
\label{55}
\end{equation}
\noindent
and below we prove that indeed it generates the correct pairing term.
It is well-known that a similar nearest-neighbor density-density attraction of
the form -$Vn_{\bf i}n_{{\bf i}+\mu}$
leads to $d$-wave superconductivity in a mean-field treatment of the one-band
repulsive $U$ Hubbard model,\cite{nazarenko} and here we merely generalize this
concept to two orbitals.

Let us discuss the mean-field treatment of $H_{\rm attr}$.
In momentum space,
the Fourier transformed of this effective attraction is:
\begin{equation}
H_{\rm attr}=-\sum_{{\bf k,k'},\alpha,\sigma}V_{\bf k,k'}d^{\dagger}_{{\bf k},\alpha,\sigma}
d^{\dagger}_{{\bf -k},-\alpha,-\sigma}d_{{\bf -k'},-\alpha,-\sigma}d_{{\bf k'},\alpha,\sigma},
\label{56}
\end{equation}
\noindent where we have requested that the pairing occurs between electrons
with opposite momentum (thus, we have dropped a third sum over all wavevectors),  
in different orbitals, and with opposite spin, as required
by the dominant singlet pairing operator obtained from the numerical simulations.
The potential is given by
\begin{eqnarray*}
V_{\bf k,k'}&=&-2V[\cos(k'_x-k_x)+\cos(k'_y-k_y)] \\
&=&V({\bf{k'-k}})=
\sum_i\tilde V_i
\eta_i({\bf{k}})\eta_i({\bf{k'}}),
\end{eqnarray*}
\noindent where $\eta_i({\bf{k}})$ are the irreducible representations of the
group $D_{\rm 4h}$. Since the ED numerical results indicate that the pairing
operator is proportional to $(\cos k_x+\cos k_y)$, which corresponds to the
irreducible representation $A_{\rm 1g}$, we will focus on that particular
term in the expansion of the full potential $V_{\bf k,k'}$.\cite{norman}
Thus, we will consider
Eq.~(\ref{56}) but using the long-range separable potential
\begin{equation}
V_{\bf k,k'}=V^* (\cos k_x+\cos k_y)(\cos k'_x+\cos k'_y),
\label{58}
\end{equation}
instead of the full short-range potential.
 
We will treat the four-fermion term in $H_{\rm attr}$ within
the usual mean-field
approximation,\cite{Bob} where some pairs of fermionic operators
are replaced by numbers, such
as $\langle b^{\dagger}_{{\bf k},\alpha}\rangle$, to be
found self-consistently. Then
$$
H_{\rm \rm MF}=H_{\rm TB}+\sum_{{\bf k,k'},\alpha}(V_{\bf k,k'}
\langle b^{\dagger}_{{\bf k},\alpha}\rangle
d_{{\bf -k'},-\alpha,\downarrow}d_{{\bf k'},\alpha,\uparrow}+
$$
\begin{equation}
V_{\bf k,k'}\langle b_{{\bf k'},\alpha}\rangle d^{\dagger}_{{\bf k},
\alpha,\uparrow} d^{\dagger}_{{\bf -k},-\alpha,\downarrow})
-\sum_{{\bf k,k'},\alpha}V_{\bf k,k'}\langle b^{\dagger}_{{\bf k},\alpha}\rangle \langle b_{{\bf k'},\alpha}\rangle.
\label{59}
\end{equation}
Defining
\begin{eqnarray}
\Delta({\bf k})&=&-\sum_{{\bf k'},\alpha}V_{\bf k,k'}\langle b_{{\bf k'},\alpha}\rangle, \nonumber \\
\Delta^{\dagger}({\bf k})&=&-\sum_{{\bf k'},\alpha}V_{\bf k,k'}\langle
b^{\dagger}_{{\bf k'},\alpha}\rangle,
\end{eqnarray}
\noindent and using the separability of the potential Eq.~(\ref{58})
we obtain
\begin{eqnarray}
H_{\rm MF}=H_{\rm TB}
+\sum_{{\bf k},{\alpha}}(\Delta^{\dagger}({\bf k})d_{{\bf -k},-\alpha,
\downarrow}d_{{\bf k},\alpha,\uparrow}+ \\
\Delta({\bf k})d^{\dagger}_{{\bf k},
\alpha,\uparrow}d^{\dagger}_{{\bf -k},-\alpha,\downarrow})
-\sum_{{\bf k,k'},\alpha}V_{\bf k,k'}\langle b^{\dagger}_{{\bf k},\alpha}\rangle \langle b_{{\bf k'},\alpha}\rangle,
\label{61}
\end{eqnarray}
\noindent which leads to the same self-consistent equations as in Section V
by setting
\begin{equation}
\Delta^{\dagger}({\bf k})=\Delta({\bf k})=V^*\Delta(\cos k_x+\cos k_y),
\end{equation}
\noindent and $V=V^*\Delta$,
where $\Delta$ should be obtained by solving the gap equation that
is obtained from minimizing
the energy of the mean-field Hamiltonian with respect
to $\Delta({\bf k})$.
 
\section{$s\pm$ pairing involving $d_{\rm xz}$ 
and $d_{\rm yz}$ electrons}
\label{app:spm}

In Section IV.B.2 we showed that our numerical simulations favored a spin
singlet
interorbital pairing state with $B_{\rm 2g}$ symmetry in the physical regime of
parameters of the two orbital model, while a pairing state with symmetry
$A_{\rm 1g}$ prevailed only in the
unphysical strong coupling regime and at the singular point $U=U'$, $J=0$.
In this appendix we will discuss in more detail the pairing operators with
$A_{\rm 1g}$ symmetry in the context of the two orbital model, and we will
show that a $s\pm$ pairing operator\cite{korshunov,kuroki,mazin,parker} 
involving only $d_{\rm xz}$ and $d_{\rm yz}$ electrons belongs to this group.

The extensive literature on the $s\pm$ pairing
state\cite{korshunov,kuroki,mazin,parker}
indicates that
this state does not have nodes on the Fermi surface and that
\begin{equation}
\Delta^1({\bf k})=-\Delta^2({\bf k+q}).
\label{smm}
\end{equation}
\noindent where 1(2) denotes the electron
(hole) Fermi surface and ${\bf q}=(\pi,0)$ or $(0,\pi)$.\cite{korshunov}
  
From the classification of possible pairing states for the
$d_{\rm xz}$ and $d_{\rm yz}$ orbitals provided in Ref.~\onlinecite{wan} we realize
that there exist the following four nodeless pairing operators: 
(i) pairing state \#1
which produces on-site intraband pairs which are even under
orbital exchange, spin singlets, and transforms according to 
the $A_{\rm 1g}$ irreducible
representation of $D_{\rm 4h}$.
Following the steps of Section V.A for this pairing state we obtain:
 
\begin{equation}
H_{\rm MF}=
 \left(\begin{array}{cccc}
\xi_{xx}  & \xi_{xy}               &  \Delta_0     & 0 \\
\xi_{xy}  & \xi_{yy}               &    0          & \Delta_0 \\
\Delta_0         & 0               & -\xi_{xx}     & -\xi_{xy}  \\
0  & \Delta_0                      & -\xi_{xy}     & -\xi_{yy}
\end{array} \right),
\label{200}
\end{equation}
\noindent where $\Delta_0$ is a constant independent of momentum.
In the base in which the tight binding Hamiltonian is diagonal we obtain:
\begin{equation}
H'_{\rm MF}=
\left(\begin{array}{cccc}
\epsilon_1 & 0               &     0  & \Delta_0 \\
0 & \epsilon_2               &     \Delta_0  &  0 \\
0  & \Delta_0 & -\epsilon_2 & 0  \\
\Delta_0  & 0 & 0 & -\epsilon_1
\end{array} \right).
\label{201}
\end{equation}
This leads to intraband pairing $\Delta_0=V$ which is momentum independent and
equal for the two bands. This does not correspond to the $s\pm$ pairing
since it does not satisfy Eq.~(\ref{smm}).\cite{korshunov}
 
Now let us consider nearest-neighbor pairing. We find that the only nodeless
pairing operators involving electrons in nearest-neighbor sites also have
symmetry $A_{\rm 1g}$ and result from a (ii) symmetric
(or (iii) antisymmetric) combination of pairings \#2 and \#3. The symmetric
(antisymmetric) combination corresponds to pairing of the  $d_{\rm xz}$ electrons
along the $x$ ($y$) direction while the $d_{\rm yz}$ pairs along the $y$ ($x$) direction.
 Following the steps of Section V.A we obtain:
 
\begin{equation}
H_{\rm MF}=
 \left(\begin{array}{cccc}
\xi_{xx}  & \xi_{xy}               &  \Delta_1     & 0 \\
\xi_{xy}  & \xi_{yy}               &    0          & \Delta_2 \\
\Delta_1         & 0               & -\xi_{xx}     & -\xi_{xy}  \\
0  & \Delta_2                      & -\xi_{xy}     & -\xi_{yy}
\end{array} \right) ,
\label{100}
\end{equation}
where $\Delta_1=V \cos k_x$ ($V \cos k_y$) and $\Delta_2=V \cos k_y$ ($V \cos k_x$)
for the symmetric (antisymmetric) combination.
In the base in which the tight binding Hamiltonian is diagonal we obtain:
 
\begin{equation}
H'_{\rm MF}=
\left(\begin{array}{cccc}
\epsilon_1 & 0               &     V_{ 12}  & V_{ 1} \\
0 & \epsilon_2               &     V_{ 2}  & V_{ 12} \\
V_{ 12}  & V_{2} & -\epsilon_2 & 0  \\
V_{ 1}  & V_{ 12} & 0 & -\epsilon_1
\end{array} \right),
\label{101}
\end{equation}
\noindent where $V_{ 1}$, $V_2$ and $V_{12}$ are given by:
\begin{equation}
V_{1}=u^2\Delta_1+v^2\Delta_2, \\
\label{102}
\end{equation}
\begin{equation}
V_{2}=v^2\Delta_1+u^2\Delta_2, \\
\label{103}
\end{equation}
\begin{equation}
V_{12}=uv(\Delta_1-\Delta_2). \\
\label{104}
\end{equation}
Thus, this leads to intraband interactions $V_1$ and $V_2$ which, according to 
our numerical checks, satisfy
$V_i({\bf k})=-V_i({\bf k+q})$ as expected for the $s\pm$ pairing,\cite{korshunov} but
there is interband pairing given by $V_{12}$ which is considered
unphysical by many authors.\cite{korshunov,kuroki,mazin,parker}
 
(iv) Finally, we can also focus on pairs of electrons along the diagonals of the square
lattice formed by the Fe ions. Following the notation of Ref.~\onlinecite{wan}
the corresponding basis function is $\cos k_x \cos k_y$ that transforms according
to $A_{\rm 1g}$. This basis provides a pairing operator with a full gap and
which is a spin singlet. It is the analog of pairing\#2
in Ref.~\onlinecite{wan} replacing the basis function $(\cos k_x+\cos k_y)$, which
represents nearest-neighbor pairing and transforms according to $A_{\rm 1g}$, by
$\cos k_x \cos k_y$ which corresponds to diagonal pairing and has the same
symmetry. We will call this pairing \#2'. It transforms
according to $A_{\rm 1g}$ and it corresponds to intraorbital pairing.
 Following the previous steps we find that for pairing \#2':
\begin{equation}
H_{\rm MF}=
 \left(\begin{array}{cccc}
\xi_{xx}  & \xi_{xy}               &  \Delta_k     & 0 \\
\xi_{xy}  & \xi_{yy}               &    0          & \Delta_k \\
\Delta_k         & 0               & -\xi_{xx}     & -\xi_{xy}  \\
0  & \Delta_k                      & -\xi_{xy}     & -\xi_{yy}
\end{array} \right) ,
\label{100a}
\end{equation}
where $\Delta_k=V \cos k_x \cos k_y$, which is exactly 
the form of the $s\pm$ pairing interaction proposed in
Ref.~\onlinecite{korshunov}.

In the base in which the tight binding Hamiltonian is diagonal we obtain:
 
\begin{equation}
H'_{\rm MF}=
\left(\begin{array}{cccc}
\epsilon_1 & 0               &     0  & \Delta_k \\
0 & \epsilon_2               &     \Delta_k  & 0 \\
0  & \Delta_k & -\epsilon_2 & 0  \\
\Delta_k  & 0 & 0 & -\epsilon_1
\end{array} \right).
\label{111}
\end{equation}

Note that this pairing operator corresponds to the $s\pm$ pairing
since it satisfies $V_i({\bf k})=-V_i({\bf k+q})$ and there is no 
interband
pairing. Thus, we have found that the $s\pm$ pairing operator is possible in
the two orbital model and transforms according to $A_{\rm 1g}$. However, the
Lanczos numerical calculations presented in the text suggest that in the
region of physical interest the undoped ground state has to be connected to
the ground state with two extra electrons via a pairing operator that
transforms according to $B_{\rm 2g}$ and, for this reason, the $s\pm$ state is not
favored. It only can prevail in the strong coupling regime of $U$ and at the
unphysical singular point $U=U'$, $J=0$ where the ground states in the doped 
and undoped regimes have the same symmetry and are connected by a pairing 
operator with $A_{\rm 1g}$ symmetry.

For completeness, let us also consider a $s\pm$ pairing operator frequently used in
the literature.\cite{korshunov,kuroki,mazin,parker} In this context
it is assumed that $\Delta^1=-\Delta^2=\Delta_0$ which is independent of the
momentum. Let us find whether this result is
consistent with the symmetry of the two-orbital model.
We start with $H'_{\rm MF}$ and working backwards the form of
$H_{\rm MF}$ is found. By this procedure we obtain
\begin{equation}
H'_{\rm MF}=
\left(\begin{array}{cccc}
\epsilon_1 & 0               &     0  & \Delta_0 \\
0 & \epsilon_2               &     -\Delta_0  & 0 \\
0  & -\Delta_0 & -\epsilon_2 & 0  \\
\Delta_0  & 0 & 0 & -\epsilon_1
\end{array} \right),
\label{105}
\end{equation}
\noindent that in terms of the original two orbitals corresponds to:
\begin{equation}
H_{\rm MF}=
 \left(\begin{array}{cccc}
\xi_{xx}  & \xi_{xy}               &  \Delta_{\rm x}     & \Delta_{\rm xy} \\
\xi_{xy}  & \xi_{yy}               & \Delta_{\rm xy}     & \Delta_{\rm y} \\
\Delta_{\rm x} & \Delta_{\rm xy}               & -\xi_{xx}     & -\xi_{xy}  \\
\Delta_{\rm xy}  & \Delta_{\rm y}  & -\xi_{xy}     & -\xi_{yy}
\end{array} \right),
\label{106}
\end{equation}
\noindent where
\begin{equation}
\Delta_{\rm x}=(u^2-v^2)\Delta_0, \\
\label{107}
\end{equation}
\begin{equation}
\Delta_{\rm y}=-(u^2-v^2)\Delta_0, \\
\label{108}
\end{equation}
\begin{equation}
\Delta_{\rm xy}=2uv\Delta_0. \\
\label{109}
\end{equation}

We have found that $uv$ transforms
according to $B_{\rm 1g}$ and $v^2-u^2$ according to $B_{\rm 2g}$, then this case
corresponds to a linear combination of on-site intraorbital (\#5) and
interorbital  (\#8) pairings that transform according to two different
irreducible representations of $D_{\rm 4h}$, i.e. $B_{\rm 1g}$ and $B_{\rm 2g}$.
The coexistance of pairs with
different symmetries can occur only if the ground state with $N$ particles
and/or the ground state with $N+2$ electrons are/is degenerate or nearly
degenerate. Numerically, we have found that the ground states appear to be
singlets and connected by an operator with symmetry $B_{\rm 2g}$ in the region of
physical relevance or $A_{\rm 1g}$ in the strong coupling limit. Only in the
unphysical singular point $J=0$ and $U=U'$ states transforming according
to $A_{\rm 1g}$, $B_{\rm 1g}$, and $B_{\rm 2g}$ are very close to each other in energy.
 
Summarizing, here it was shown that the $s\pm$ pairing operator that pairs
electrons
along the diagonals of the Fe square lattice using
the $d_{\rm xz}$ and $d_{\rm yz}$ orbitals does transform according
to the $A_{\rm 1g}$ irreducible representation of the $D_{\rm 4h}$ group that
characterizes the symmetry of the Fe-As planes in the pnictides. Numerically we
found that this is the pairing symmetry that prevails in the strong coupling
region
but that in the physical regime the pairing operator must transform according
to $B_{\rm 2g}$. In addition, we have observed that the often-used 
momentum-independent approximation
for the $s\pm$ operator does not respect the
symmetry of the Fe As planes.


\begin{thebibliography}{100}

\bibitem{Fe-SC}
Y. Kamihara, T. Watanabe, M. Hirano, and H. Hosono, J. of the Am. Chem. Soc.
  {\bf 130},  3296  (2008).

\bibitem{chen1}
G.~F. Chen, Z. Li, G. Li, J. Zhou, D. Wu, J. Dong, W.~Z. Hu, P. Zheng, Z.~J.
  Chen, H.~Q. Yuan, J. Singleton, J.~L. Luo, and N.~L. Wang, Phys. Rev. Lett.
  {\bf 101},  057007  (2008).

\bibitem{chen2}
G.~F. Chen, Z. Li, D. Wu, G. Li, W.~Z. Hu, J. Dong, P. Zheng, J.~L. Luo, and
  N.~L. Wang, Phys. Rev. Lett. {\bf 100},  247002  (2008).

\bibitem{wen}
H.-H. Wen, G. Mu, L. Fang, H. Yang, and X. Zhu, EPL {\bf 82},  17009  (2008).

\bibitem{chen3}
{X. H. Chen}, {T. Wu}, {G. Wu}, {R. H. Liu}, {H. Chen}, and {D. F. Fang},
  Nature {\bf 453},  761  (2008).

\bibitem{ren1}
{Ren, Z.A.}, {Yang, J.}, {Lu, W.}, {Yi, W.}, {Che, G.C.}, {Dong, X.L.}, {Sun,
  L.L.}, and {Zhao, Z.X.}, Mater. Res. Innovat. {\bf 12},  105  (2008).

\bibitem{55}
{Ren Zhi-An}, {Lu Wei}, {Yang Jie}, {Yi Wei}, {Shen Xiao-Li}, {Zheng-Cai}, {Che
  Guang-Can}, {Dong Xiao-Li}, {Sun Li-Ling}, {Zhou Fang}, and {Zhao
  Zhong-Xian}, Chin. Phys. Lett. {\bf 25},  2215  (2008).

\bibitem{ren2}
Z.-A. Ren, G.-C. Che, X.-L. Dong, J. Yang, W. Lu, W. Yi, X.-L. Shen,
Z.-C. Li, L.-L. Sun, F. Zhou, and Z.-X. Zhao, EPL {\bf 83}, 17002 (2008).

\bibitem{phonon0}
L. Boeri, O.~V. Dolgov, and A.~A. Golubov, Phys. Rev. Lett. {\bf 101},  026403
  (2008).

\bibitem{phonon1}
S. Higashitaniguchi, M. Seto, S. Kitao, Y. Kobayashi, M. Saito, R. Masuda, T.
  Mitsui, Y. Yoda, Y. Kamihara, M. Hirano, and H. Hosono, Phys. Rev. B
  {\bf 78}, 174507 (2008).

\bibitem{phonon2}
A.~D. Christianson, M.~D. Lumsden, O. Delaire, M.~B. Stone, D.~L. Abernathy,
  M.~A. McGuire, A.~S. Sefat, R. Jin, B.~C. Sales, D. Mandrus, E.~D. Mun, P.~C.
  Canfield, J.~Y.~Y. Lin, M. Lucas, M. Kresch, J.~B. Keith, B. Fultz, E.~A.
  Goremychkin, and R.~J. McQueeney, Phys. Rev. Lett. {\bf 101},  157004
  (2008).

\bibitem{sefat}
A.~S. Sefat, M.~A. McGuire, B.~C. Sales, R. Jin, J.~Y. Howe, and D. Mandrus,
  Phys. Rev. B {\bf 77},  174503  (2008).

\bibitem{liu}
R.~H. Liu, G. Wu, T. Wu, D.~F. Fang, H. Chen, S.~Y. Li, K. Liu, Y.~L. Xie,
  X.~F. Wang, R.~L. Yang, L. Ding, C. He, D.~L. Feng, and X.~H. Chen, Phys.
  Rev. Lett. {\bf 101},  087001  (2008).

\bibitem{haule2}
K. Haule, J.~H. Shim, and G. Kotliar, Phys. Rev. Lett. {\bf 100},  226402
  (2008).

\bibitem{haule}
K. Haule and G. Kotliar, arXiv:0805.0722, 2008.

\bibitem{dubroka}
A. Dubroka, K.~W. Kim, M. R\"{o}ssle, V.~K. Malik, A.~J. Drew, R.~H. Liu, G.
  Wu, X.~H. Chen, and C. Bernhard, Phys. Rev. Lett. {\bf 101},  097011  (2008).

\bibitem{boris}
A.~V. Boris, N.~N. Kovaleva, S.~S.~A. Seo, J.~S. Kim, P. Popovich, Y. Matiks,
  R.~K. Kremer, and B. Keimer, arXiv:0806.1732, 2008.

\bibitem{C.Liu}
C. Liu, T. Kondo, M.~E. Tillman, R. Gordon, G.~D. Samolyuk, Y. Lee, C. Martin,
  J.~L. McChesney, S. Bud'ko, M.~A. Tanatar, E. Rotenberg, P.~C. Canfield, R.
  Prozorov, B.~N. Harmon, and A. Kaminski, arXiv:0806.2147, 2008.

\bibitem{J.Zhao}
{Jun Zhao}, {Q. Huang}, {C. de la Cruz}, {S. Li}, {J. W. Lynn}, {Y. Chen}, {M.
  A. Green}, {G. F. Chen}, {G. Li}, {Z. Li}, {J. L. Luo}, {N. L. Wang}, and {P.
  Dai}, Nat. Mater. {\bf 7},  953  (2008).

\bibitem{kohama}
Y. Kohama, Y. Kamihara, H. Kawaji, T. Atake, M. Hirano, and H. Hosono, J. Phys.
  Soc. Jpn. {\bf 77},  094715  (2008).

\bibitem{imada}
K. Nakamura, R. Arita, and M. Imada, J. Phys. Soc. Jpn. {\bf 77},  093711
  (2008).

\bibitem{H.Liu}
H. Liu, W. Zhang, L. Zhao, X. Jia, J. Meng, G. Liu, X. Dong, G.~F. Chen, J.~L.
  Luo, N.~L. Wang, W. Lu, G. Wang, Y. Zhou, Y. Zhu, X. Wang, Z. Xu, C. Chen,
  and X.~J. Zhou, Phys. Rev. B {\bf 78},  184514  (2008).

\bibitem{Jaro}
J. Jaroszynski, S.~C. Riggs, F. Hunte, A. Gurevich, D.~C. Larbalestier, G.~S.
  Boebinger, F.~F. Balakirev, A. Migliori, Z.~A. Ren, W. Lu, J. Yang, X.~L.
  Shen, X.~L. Dong, Z.~X. Zhao, R. Jin, A.~S. Sefat, M.~A. McGuire, B.~C.
  Sales, D.~K. Christen, and D. Mandrus, Phys. Rev. B {\bf 78},  064511
  (2008).

\bibitem{basov}
M.~M. Qazilbash, J.~J. Hamlin, R.~E. Baumbach, M.~B. Maple, and D.~N. Basov,
  arXiv:0808.3748, 2008.

\bibitem{Y.Ishida}
Y. Ishida, T. Shimojima, K. Ishizaka, T. Kiss, M. Okawa, T. Togashi, S.
  Watanabe, X.~Y. Wang, C.~T. Chen, Y. Kamihara, M. Hirano, H. Hosono, and S.
  Shin, arXiv:0805.2647, 2008.

\bibitem{T.Sato}
T. Sato, S. Souma, K. Nakayama, K. Terashima, K. Sugawara, T. Takahashi, Y.
  Kamihara, M. Hirano, and H. Hosono, J. Phys. Soc. Jpn. {\bf 77},  063708
  (2008).

\bibitem{liu2}
L. Hai-Yun, J. Xiao-Wen, Z. Wen-Tao, Z. Lin, M. Jian-Qiao, L. Guo-Dong, D.
  Xiao-Li, W. Gang, L. Rong-Hua, C. Xian-Hui, R. Zhi-An, Y. Wei, C. Guang-Can,
  C. Gen-Fu, W. Nan-Lin, W. Gui-Ling, Z. Yong, Z. Yong, W. Xiao-Yang, Z.
  Zhong-Xian, X. Zu-Yan, C. Chuang-Tian, and Z. Xing-Jiang, Chin. Phys. Lett.
  {\bf 25},  3761  (2008).

\bibitem{L.Zhao}
L. Zhao, H. Liu, W. Zhang, J. Meng, X. Jia, G. Liu, X. Dong, G.~F. Chen, J.~L.
  Luo, N.~L. Wang, G. Wang, Y. Zhou, Y. Zhu, X. Wang, Z. Zhao, Z. XU, C. Chen,
  and X.~J. Zhou, Chin. Phys. Lett. {\bf 25},  4402  (2008).

\bibitem{A.Drew}
A.~J. Drew, F.~L. Pratt, T. Lancaster, S.~J. Blundell, P.~J. Baker, R.~H. Liu,
  G. Wu, X.~H. Chen, I. Watanabe, V.~K. Malik, A. Dubroka, K.~W. Kim, M.
  R\"{o}ssle, and C. Bernhard, Phys. Rev. Lett. {\bf 101},  097010  (2008).

\bibitem{I.Felner}
I. Felner, I. Nowik, M.~I. Tsindlekht, Z.-A. Ren, X.-L. Shen, G.-C. Che, and
  Z.-X. Zhao, arXiv:0805.2794, 2008.

\bibitem{takeshita}
S. Takeshita, R. Kadono, M. Hiraishi, M. Miyazaki, A. Koda, Y. Kamihara, and H.
  Hosono, J. Phys. Soc. Jpn. {\bf 77},  103703  (2008).

\bibitem{H.Chen}
H. Chen, Y. Ren, Y. Qiu, W. bao, R.~H. Liu, G. Wu, T. Wu, Y.~L. Xie, X.~F.
  Wang, Q. Huang, and X.~H. Chen, EPL {\bf 85}, 17006 (2009)

\bibitem{nodal1}
{Shan, Lei }, {Wang, Yonglei }, {Zhu, Xiyu }, {Mu, Gang }, {Fang, Lei }, {Ren,
  Cong }, and {Wen, Hai-Hu }, EPL {\bf 83},  57004  (2008).

\bibitem{nodal2}
M. Gang, Z. Xi-Yu, F. Lei, S. Lei, R. Cong, and W. Hai-Hu, Chin. Phys. Lett.
  {\bf 25},  2221  (2008).

\bibitem{nodal3}
C. Ren, Z.-S. Wang, H. Yang, X. Zhu, L. Fang, G. Mu, L. Shan, and H.-H. Wen,
  arXiv:0804.1726, 2008.

\bibitem{Ahilan}
K. Ahilan, F.~L. Ning, T. Imai, A.~S. Sefat, R. Jin, M.~A. McGuire, B.~C.
  Sales, and D. Mandrus, Phys. Rev. B {\bf 78},  100501  (2008).

\bibitem{nakai}
Y. Nakai, K. Ishida, Y. Kamihara, M. Hirano, and H. Hosono, J. Phys. Soc. Jpn.
  {\bf 77},  073701  (2008).

\bibitem{Grafe}
H.-J. Grafe, D. Paar, G. Lang, N.~J. Curro, G. Behr, J. Werner, J.
  Hamann-Borrero, C. Hess, N. Leps, R. Klingeler, and B. B\"{u}chner, Phys.
  Rev. Lett. {\bf 101},  047003  (2008).

\bibitem{Y.Wang}
Y.-L. Wang, L. Shan, L. Fang, P. Cheng, C. Ren, and H.-H. Wen, Supercond. Sci.
  Technol. {\bf 22},  015018  (2009).

\bibitem{matano}
{Matano, K. }, {Ren, Z. A.}, {Dong, X. L.}, {Sun, L. L.}, {Zhao, Z. X.}, and
  {Zheng, Guo-qing }, EPL {\bf 83},  57001  (2008).

\bibitem{mukuda}
H. Mukuda, N. Terasaki, H. Kinouchi, M. Yashima, Y. Kitaoka, S. Suzuki, S.
  Miyasaka, S. Tajima, K. Miyazawa, P. Shirage, H. Kito, H. Eisaki, and A. Iyo,
  J. Phys. Soc. Jpn. {\bf 77},  093704  (2008).

\bibitem{millo}
O. Millo, I. Asulin, O. Yuli, I. Felner, Z.-A. Ren, X.-L. Shen, G.-C. Che, and
  Z.-X. Zhao, Phys. Rev. B {\bf 78},  092505  (2008).

\bibitem{wang-nodes}
X.~L. Wang, S.~X. Dou, Z.-A. Ren, W. Yi, Z.-C. Li, Z.-X. Zhao, and S.-I. Lee,
  arXiv:0808.3398, 2008.

\bibitem{hashimoto}
K. Hashimoto, T. Shibauchi, T. Kato, K. Ikada, R. Okazaki, H. Shishido, M.
  Ishikado, H. Kito, A. Iyo, H. Eisaki, S. Shamoto, and Y. Matsuda, Phys. Rev. Lett. {\bf 102}, 017002 (2009).

\bibitem{arpes}
T. Kondo, A.~F. Santander-Syro, O. Copie, C. Liu, M.~E. Tillman, E.~D. Mun, J.
  Schmalian, S.~L. Bud'ko, M.~A. Tanatar, P.~C. Canfield, and A. Kaminski,
  Phys. Rev. Lett. {\bf 101},  147003  (2008).

\bibitem{arpes2}
H. Ding, P. Richard, K. Nakayama, K. Sugawara, T. Arakane, Y. Sekiba, A.
  Takayama, S. Souma, T. Sato, T. Takahashi, Z. Wang, X. Dai, Z. Fang, G.~F.
  Chen, J.~L. Luo, and N.~L. Wang, EPL {\bf 83},  47001  (2008).

\bibitem{C.Martin}
C. Martin, R.~T. Gordon, M.~A. Tanatar, M.~D. Vannette, M.~E. Tillman, E.~D.
  Mun, P.~C. Canfield, V.~G. Kogan, G.~D. Samolyuk, J. Schmalian, and R.
  Prozorov, arXiv:0807.0876, 2008.

\bibitem{T.Chen}
T.~Y. Chen, Z. Tesanovic, R.~H. Liu, X.~H. Chen, and C.~L. Chien, Nature {\bf
  453},  1224  (2008).

\bibitem{parker}
D. Parker, O.~V. Dolgov, M.~M. Korshunov, A.~A. Golubov, and I.~I. Mazin, Phys.
  Rev. B {\bf 78},  134524  (2008).

\bibitem{mu}
G. Mu, H. Luo, Z. Wang, L. Shan, C. Ren, and H.-H. Wen, arXiv:0808.2941, 2008.

\bibitem{sdw}
J. Dong, H.~J. Zhang, G. Xu, Z. Li, G. Li, W.~Z. Hu, D. Wu, G.~F. Chen, X. Dai,
  J.~L. Luo, Z. Fang, and N.~L. Wang, EPL {\bf 83},  27006  (2008).

\bibitem{neutrons1}
{C. de la Cruz}, {Q. Huang}, {J. W. Lynn}, {Jiying Li}, {W. Ratcliff II}, {J.
  L. Zarestky}, {H. A. Mook}, {G. F. Chen}, {J. L. Luo}, {N. L. Wang}, and {P.
  Dai}, Nature {\bf 453},  899  (2008).

\bibitem{neutrons2}
Y. Chen, J.~W. Lynn, J. Li, G. Li, G.~F. Chen, J.~L. Luo, N.~L. Wang, P. Dai,
  C. dela Cruz, and H.~A. Mook, Phys. Rev. B {\bf 78},  064515  (2008).

\bibitem{neutrons3}
C. Krellner, N. Caroca-Canales, A. Jesche, H. Rosner, A. Ormeci, and C. Geibel,
  Phys. Rev. B {\bf 78},  100504  (2008).

\bibitem{neutrons4}
A.~I. Goldman, D.~N. Argyriou, B. Ouladdiaf, T. Chatterji, A. Kreyssig, S.
  Nandi, N. Ni, S.~L. Bud'ko, P.~C. Canfield, and R.~J. McQueeney, Phys. Rev. B
  {\bf 78},  100506  (2008).

\bibitem{first}
S. Lebegue, Phys. Rev. B {\bf 75},  035110  (2007).

\bibitem{singh}
D.~J. Singh and M.-H. Du, Phys. Rev. Lett. {\bf 100},  237003  (2008).

\bibitem{xu}
G. Xu, W. Ming, Y. Yao, X. Dai, S.-C. Zhang, and Z. Fang, EPL {\bf 82},  67002
  (2008).

\bibitem{cao}
C. Cao, P.~J. Hirschfeld, and H.-P. Cheng, Phys. Rev. B {\bf 77},  220506
  (2008).

\bibitem{fang2}
H.-J. Zhang, G. Xu, X. Dai, and Z. Fang, Chin. Phys. Lett. {\bf 26}, 017401 (2009).

\bibitem{kuroki}
K. Kuroki, S. Onari, R. Arita, H. Usui, Y. Tanaka, H. Kontani, and H. Aoki,
  Phys. Rev. Lett. {\bf 101},  087004  (2008).

\bibitem{mazin}
I.~I. Mazin, D.~J. Singh, M.~D. Johannes, and M.~H. Du, Phys. Rev. Lett. {\bf
  101},  057003  (2008).

\bibitem{FCZhang}
X. Dai, Z. Fang, Y. Zhou, and F.-C. Zhang, Phys. Rev. Lett. {\bf 101},  057008
  (2008).

\bibitem{han}
Q. Han, Y. Chen, and Z.~D. Wang, EPL {\bf 82},  37007  (2008).

\bibitem{korshunov}
M.~M. Korshunov and I. Eremin, Phys. Rev. B {\bf 78},  140509  (2008).

\bibitem{baskaran}
G. Baskaran, arXiv:0804.1341, 2008.

\bibitem{plee}
P.~A. Lee and X.-G. Wen, arXiv:0804.1739, 2008.

\bibitem{yildirim}
T. Yildirim, Phys. Rev. Lett. {\bf 101},  057010  (2008).

\bibitem{extra}
Q. Si and E. Abrahams, Phys. Rev. Lett. {\bf 101},  076401  (2008).

\bibitem{yao}
Z.-J. Yao, J.-X. Li, and Z.~D. Wang, arXiv:0804.4166, 2008.

\bibitem{xu2}
C. Xu, M. M\"{u}ller, and S. Sachdev, Phys. Rev. B {\bf 78},  020501  (2008).

\bibitem{stratos}
E. Manousakis, J. Ren, S. Meng, and E. Kaxiras, Phys. Rev. B {\bf 78},  205112
  (2008).

\bibitem{scalapino}
S. Raghu, X.-L. Qi, C.-X. Liu, D.~J. Scalapino, and S.-C. Zhang, Phys. Rev. B
  {\bf 77},  220503  (2008).

\bibitem{li}
T. Li, J. Phys.: Condens. Matter {\bf 20},  425203 (6pp)  (2008).

\bibitem{qi}
X.-L. Qi, S. Raghu, C.-X. Liu, D.~J. Scalapino, and S.-C. Zhang,
  arXiv:0804.4332, 2008.

\bibitem{ours}
M. Daghofer, A. Moreo, J.~A. Riera, E. Arrigoni, D.~J. Scalapino, and E.
  Dagotto, Phys. Rev. Lett. {\bf 101}, 237004 (2008).

\bibitem{hu}
K. Seo, B.~A. Bernevig, and J. Hu, Phys. Rev. Lett. {\bf 101},  206404  (2008).

\bibitem{dhlee}
Y. Ran, F. Wang, H. Zhai, A. Vishwanath, and D.-H. Lee, arXiv:0805.3535, 2008.

\bibitem{zhou}
Y. Zhou, W.-Q. Chen, and F.-C. Zhang, Phys. Rev. B {\bf 78},  064514  (2008).

\bibitem{lorenzana}
J. Lorenzana, G. Seibold, C. Ortix, and M. Grilli, Phys. Rev. Lett. {\bf 101},
  186402  (2008).

\bibitem{sk}
R. Sknepnek, G. Samolyuk, Y. bin Lee, B.~N. Harmon, and J. Schmalian,
  arXiv:0807.4566, 2008.

\bibitem{parish}
M.~M. Parish, J. Hu, and B.~A. Bernevig, Phys. Rev. B {\bf 78},  144514
  (2008).

\bibitem{choi}
H.-Y. Choi and Y. Bang, arXiv:0807.4604, 2008.

\bibitem{yang}
S. Yang, W.-L. You, S.-J. Gu, and H.-Q. Lin, arXiv:0807.0587, 2008.

\bibitem{calderon}
M.~J. Calderon, B. Valenzuela, and E. Bascones, arXiv:0810.0019, 2008.

\bibitem{wang}
Z.-H. Wang, H. Tang, Z. Fang, and X. Dai, arXiv:0805.0736, 2008.

\bibitem{shi}
J. Shi, arXiv:0806.0259, 2008.

\bibitem{2orbitals}
W.-L. You, S.-J. Gu, G.-S. Tian, and H.-Q. Lin, arXiv:0807.1493, 2008.

\bibitem{wan}
Y. Wan and Q.-H. Wang, arXiv:0805.0923, 2008.

\bibitem{kawabata}
A. Kawabata, S.~C. Lee, T. Moyoshi, Y. Kobayashi, and M. Sato, arXiv:0807.3480,
  2008.

\bibitem{vero}
V. Vildosola, L. Pourovskii, R. Arita, S. Biermann, and A. Georges, Phys. Rev.
  B {\bf 78},  064518  (2008).

\bibitem{Suhl}
H. Suhl, B.~T. Matthias, and L.~R. Walker, Phys. Rev. Lett. {\bf 3},  552
  (1959).

 \bibitem{slater}
J.~C. Slater and G.~F. Koster, Phys. Rev. {\bf 94},  1498  (1954).

\bibitem{fulde}
For more details, see P. Fulde, {\it Electron Correlations in Molecules and
  Solids}, Springer Series in Solid-State Sciences 100, 1991.

\bibitem{Harrison}
W.~A. Harrison, {\em Electronic Structure and the Properties of Solids} (Dover
  Publications, New York, 1989).

\bibitem{RMP01}
E. Dagotto, T. Hotta, and A. Moreo, Phys. Rep. {\bf 344},  1   (2001).

\bibitem{Rev1994}
E. Dagotto, Rev. Mod. Phys. {\bf 66},  763  (1994).

\bibitem{comment}
For small $|pd\pi|<0.15$, we find other states. They have $p$ symmetry and
  their momentum is $(\pi,\pi)$ instead of $(0,0)$. Since these parameters
  locate us close to the phase transition of the half-filled system at $pd\pi
  \approx 0$, where the Fermi surface topology changes [see
  Fig.~\ref{fig:fs_pdp0}] and the crucial $(0,\pi)$ spin order breaks down [see
  Fig.~\ref{fig:sk_pdpi0.1}], we tentatively attribute these exotic results
  with finite-momentum pairing to finite size effects.

\bibitem{kumar} N. Kumar and K.P. Sinha, Phys.~Rev.~{\bf 174}, 482 (1968).

\bibitem{kheli} Jamil Tahir-Kheli, Phys.~Rev.~{\bf B58}, 12307 (1998).

\bibitem{khomskii}
{O.V. Dolgov}, {E.P. Fetisov}, {D.I. Khomskii}, and {K. Svozil}, Z. Phys. B
  {\bf 67},  63  (1987).

\bibitem{Yu} R. Yu, K. Trinh, A. Moreo, M. Daghofer, J. Riera, S. Haas, and 
E. Dagotto, preprint. 

\bibitem{another}
Another way to understand the ``non-trivial'' nodes in the MF solution is the
following: if $\xi_{xy}=0$, which happens if $\sin (k_i)=0$ (for $i=x$ or $y$) as
already discussed, then the original
Hamiltonian Eq.~(\ref{42}) has a simple form. The nodes will occur 
at values of $(k_x,k_y)$
where the determinant of Eq.~(\ref{42}) with $\xi_{xy}=0$ vanishes. The resulting 
equation is given by $\Delta_k^2+\xi_{xx}\xi_{yy}=0.$
Note now that $\xi_{ii}$ and $\Delta_k$ are only functions of $\cos k_x$ or $\cos k_y$,
since one of these cosines is fixed to 1 or -1 due to 
$\xi_{xy}=0$. The product $\xi_{xx} \xi_{yy}$ has to
be negative indicating that, if the resulting quadratic equation has real solutions, the
nodes will appear at points that lie in
between the two non-interacting Fermi surfaces, as derived before.


\bibitem{arpes22}
L. Wray, D. Qian, D. Hsieh, Y. Xia, L. Li, J.~G. Checkelsky, A. Pasupathy,
  K.~K. Gomes, A.~V. Fedorov, G.~F. Chen, J.~L. Luo, A. Yazdani, N.~P. Ong,
  N.~L. Wang, and M.~Z. Hasan, arXiv:0808.2185, 2008. In
  Ref.~\onlinecite{C.Liu} it was also concluded that the regime of strong
  coupling was realized in the Fe pnictides.

\bibitem{arpes21}
A similar conclusion regarding the importance of further analyzing the electron
  pockets holds for the ARPES study reported in Ref.~\onlinecite{L.Zhao}.

\bibitem{arpes3}
V.~B. Zabolotnyy, D.~S. Inosov, D.~V. Evtushinsky, A. Koitzsch, A.~A. Kordyuk,
  J.~T. Park, D. Haug, V. Hinkov, A.~V. Boris, D.~L. Sun, G.~L. Sun, C.~T. Lin,
  B. Keimer, M. Knupfer, B. Buechner, A. Varykhalov, R. Follath, and S.~V.
  Borisenko, arXiv:0808.2454, 2008.

\bibitem{foot} We have observed that pairing $\#8$ ($B_{\rm 2g}$ on site) that, as
mentioned in the text, appears to coexist with pairing $\#9$ ($B_{\rm 2g}$ nearest
neighbors), produces a nodeless gap for some values of $V$. However, only 
combinations of pairing $\#8$ and $\#9$  in which $V$ is stronger for $\#8$
provide a nodeless gap while the numerical results appear to indicate that $V$
should be smaller for $\#8$.

\bibitem{nazarenko}
See for instance E. Dagotto, J. Riera, Y. C. Chen, A. Moreo, A. Nazarenko, F.
  Alcaraz, and F. Ortolani, Phys. Rev. B {\bf 49}, 3548 (1994); Alexander
  Nazarenko, Adriana Moreo, Elbio Dagotto, and Jose Riera, Phys. Rev. B {\bf
  54}, R768 (1996), and references therein.

\bibitem{norman}
R. Fehrenbacher and M.~R. Norman, Phys. Rev. Lett. {\bf 74},  3884  (1995).

\bibitem{Bob}
J. Schrieffer, {\em Theory of Superconductivity}, 4 ed. (Addison-Wesley, New
  York, 1988).

\end{thebibliography}

\end{document}